\documentclass[useAMS,usenatbib]{mn2e}

\usepackage{amsmath}
\usepackage{amssymb} 
\usepackage{graphicx} 
\usepackage{comment}
 
\newcommand{\qB}{q^\mathrm{B}}
\newcommand{\qS}{q^\mathrm{S}}
\newcommand{\qBS}{q^\mathrm{BS}}

\newcommand{\lddp}[2]{\upartial #1/\upartial #2}

\newcommand{\sech}{\mathrm{sech}}
\newcommand{\potb}{\Phi_{\mathrm{b}}}

\newcommand{\Rb}{R_{\mathrm{b}}}
\newcommand{\Rd}{h_{\mathrm{R}}}
\newcommand{\Rs}{R_{\mathrm{s}}}
\newcommand{\Rsg}{R_\sigma}
\newcommand{\Omegab}{\Omega_\mathrm{b}}
\newcommand{\Omegas}{\Omega_\mathrm{s}}

\newcommand{\ROLR}{R_{\mathrm{OLR}}}

\newcommand{\Km}{~\mathrm{km}}
 
\newcommand{\Kpc}{~\mathrm{kpc}}
\newcommand{\pc}{~\mathrm{pc}} 
\newcommand{\Gyr}{~\mathrm{Gyr}}

\newcommand{\kmsec}{~\mathrm{km}~\mathrm{s}^{-1}}
\newcommand{\kmsect}{~\mathrm{km^2}~\mathrm{s}^{-2}}
\newcommand{\kmseckpc}{~\mathrm{km}~\mathrm{s}^{-1}~\mathrm{kpc}^{-1}}

\newcommand{\Msun}{M_{\odot}}

\newcommand{\de}{\mathrm{d}}

\newcommand{\Rg}{R_\mathrm{g}} 
\newcommand{\Rgz}{R_{\mathrm{g},0}}

\newcommand{\RNum}[1]{\uppercase\expandafter{\romannumeral #1\relax}}

\newcommand{\ti}{t_\mathrm{i}}
\newcommand{\te}{t_\mathrm{e}}

\newcommand{\vlos}{v_\mathrm{los}}
\newcommand{\vlost}{\tilde{v}_\mathrm{los}}
\newcommand{\vlosz}{v_\mathrm{los,0}}

\newcommand{\pare}[1]{\left(#1\right)}
\newcommand{\paresq}[1]{\left[#1\right]}

\newcommand{\absp}[1]{\left|#1\right|}
\newcommand{\av}[1]{\langle #1 \rangle}
\newcommand{\avvR}{\av{v_R}}%{\langle v_R \rangle}
\newcommand{\avvphi}{\av{v_\phi}}
\newcommand{\tavvphi}{\tilde{\av{v_\phi}}}
\newcommand{\avvphiz}{\av{v_{\phi,0}}}
\newcommand{\avvz}{\av{v_z}}

\newcommand{\Eq}[1]{equation~(\ref{#1})}

\newcommand{\Fig}[1]{Fig.~\ref{#1}}
\newcommand{\Figs}[2]{Figs.~\ref{#1}-\ref{#2}}

\newcommand{\dvz}{\Delta \av{v_z}}

\newcommand{\Sec}[1]{Section~\ref{#1}}

\newcommand{\Phis}{\Phi_{\mathrm{s}}}
\newcommand{\rhos}{\rho_{\mathrm{s}}}
\newcommand{\hs}{h_{\mathrm{s}}}
\newcommand{\phis}{\phi_\mathrm{s}}
\newcommand{\phib}{\phi_\mathrm{b}}
\newcommand{\gammab}{\gamma_\mathrm{b}}
\newcommand{\gammas}{\gamma_\mathrm{s}}
\newcommand{\Pk}[1]{P_{k,\mathrm{#1}}}
\newcommand{\elos}{\mathbf{e}_\mathrm{los}}
\newcommand{\eR}{\mathbf{e}_R}
\newcommand{\ephi}{\mathbf{e}_\phi}
\newcommand{\kmax}{k_\mathrm{max}}

\title[The effects of Galactic bar-spiral coupling]{The effects
  of bar-spiral coupling on stellar kinematics in the Galaxy}
\author[G. Monari et al.]  {\parbox[t]{\textwidth}{ Giacomo
    Monari$^{1}$\thanks{giacomo.monari@astro.unistra.fr}, Benoit
    Famaey$^{1}$, Arnaud Siebert$^{1}$, Robert J.J. Grand$^{23}$,
    \\ Daisuke Kawata$^4$, Christian Boily$^{1}$}
  \vspace{10pt} \\ $^1$Observatoire astronomique de Strasbourg,
  Universit\'e de Strasbourg, CNRS, UMR 7550, 11 rue de
  l'Universit\'e, F-67000 Strasbourg, France \\ $^2$Heidelberger
  Institut f\"{u}r Theoretische Studien, Schloss-Wolfsbrunnenweg 35,
  69118 Heidelberg, Germany \\ $^3$Zentrum f\"{u}r Astronomie der
  Universit\"{a}t Heidelberg, Astronomisches Recheninstitut,
  M\"{o}nchhofstr. 12-14, 69120 Heidelberg, Germany \\ $^4$Mullard
  Space Science Laboratory, University College London, Holmbury
  St. Mary, Dorking, Surrey, RH5 6NT, United Kingdom}

\pagerange{\pageref{firstpage}--\pageref{lastpage}} \pubyear{2015}

\def\LaTeX{L\kern-.36em\raise.3ex\hbox{a}\kern-.15em
    T\kern-.1667em\lower.7ex\hbox{E}\kern-.125emX}

\begin{document}

\label{firstpage}

\maketitle

\begin{abstract}
  We investigate models of the Milky Way disc taking into account
  simultaneously the bar and a two-armed quasi-static spiral
  pattern. Away from major resonance overlaps, the mean stellar radial
  motions in the plane are essentially a linear superposition of the
  isolated effects of the bar and spirals. Thus, provided the bar is
  strong enough, even in the presence of spiral arms, these mean
  radial motions are predominantly affected by the Galactic bar for
  large scale velocity fluctuations. This is evident when comparing
  the peculiar line-of-sight velocity power spectrum of our coupled
  models with bar-only models. However, we show how forthcoming
  spectroscopic surveys could disentangle bar-only non-axisymmetric
  models of the Galaxy from models in which spiral arms have a
  significant amplitude. We also point out that overlaps of low-order
  resonances are sufficient to enhance stellar churning within the
  disc, even when the spirals amplitude is kept
  constant. Nevertheless, for churning to be truly non-local, stronger
  or (more likely) transient amplitudes would be needed: otherwise the
  disc is actually mostly unaffected by churning in the present
  models. Finally, regarding vertical breathing modes, the combined
  effect of the bar and spirals on vertical motions is a clear
  non-linear superposition of the isolated effects of both components,
  significantly superseding the linear superposition of modes produced
  by each perturber separately, thereby providing an additional effect
  to consider when analysing the observed breathing mode of the
  Galactic disc in the extended Solar neighbourhood.
\end{abstract}


\section{Introduction}

The Milky Way disc has long been known to possess non-axisymmetries,
essentially in the form of a central bar and spiral arms. But our
detailed understanding of the nature and of the dynamical effects of
these structures is still in its infancy. These structures are
important because they are significant drivers of
%likely key players in the 
dynamics and evolution of the Galaxy, through effects such as in-plane
heating and radial migration
\citep{SellwoodBinney2002,MinchevFamaey2010}.

The nature and origin of spiral arms is still a matter of debate, and
interpretations of spirals in self-consistent numerical simulations
range from very transient co-rotating structures (dynamic spirals)
which wind up and disappear over time
%winding up and disappearing with time
\citep[e.g.][]{Grand2012,Baba2013,Grand2015} to multiple long-lived
($\sim 10$ galaxy rotations) modes
\citep[e.g.][]{Minchev2012AA548A126, Quillen2011, DOnghia2013,
  SellwoodCarlberg2014}. In the latter case, even though such modes do
not appear to be strictly static as in the classical density wave
picture, they are nevertheless genuine standing wave oscillations with
fixed shape and pattern speed. In principle the response of stars and
gas to these waves away from the main resonances can be computed from
linear perturbation theory \citep[][hereafter
M16]{LinShu,LinShu1966,Monari2016} and can be simply added to each
other if there is no nonlinear density growth when the modes overlap
each other. Hence it is interesting to consider the response to single
modes in test-particle simulations to get an insight of the effects of
such modes on the kinematics of stellar populations of the Galactic
disc. Unlike \cite{Grand2015}, we will consider here non-varying amplitudes only.
Note that our models will also differ from those by not taking self-gravity
into account, but have the advantage of controlling the strength of
the perturbation. Such test-particle simulations can be very useful as
benchmarks for analytical models such as those developed in M16. Such
simulations have also allowed to demonstrate in the past how local
velocity-space substructures made of stars of different ages and
chemical compositions, known as moving groups in the Solar
neighbourhood \citep[e.g.][]{Chereul1998,Dehnen1998,Famaey2005}, are
typical responses to a given spiral mode near its resonances
\citep[e.g.][]{QuillenMinchev2005,Pompeia2011,Antoja2011}. The outer
Lindblad resonance (OLR) from the central bar is generally
acknowledged to play a similar role in explaining the kinematic group
known as the Hercules stream
\citep[e.g.][]{Dehnen2000,Monari2013}. The role of the bar in driving
or sustaining spiral arms is on the other hand still unclear. While
there are clear mechanisms for generating bar-driven spiral arms
sharing the same pattern speed as the bar
\citep[e.g.][]{RomeroGomez2007,Sormani2015}, there is also evidence
that the Milky Way central bar and the main spiral pattern at the
Solar position do not share the same pattern speed. However, even the
pattern speed of the Milky Way bar itself is still subject to debate,
as recent results from \cite{Wegg2015} and \cite{Portail2015} argue
for a much smaller pattern speed than previously estimated. This would
nevertheless make it difficult to explain the presence of the Hercules
stream in the Solar vicinity (Monari~et~al.~in preparation).

The probable combined presence of a bar and spiral arms having {\it
  different} pattern speeds in our Galaxy thus makes it of utmost
importance to understand how their combination affects stellar
kinematics. Away from major resonances, as stated above, it is a
priori expected that the average in-plane motions are a linear
superposition of both. It is nevertheless important to understand the
behavior at resonances too, both in terms of radial migration of stars
in the disc\footnote{This is referred to as `churning' when not
  accompanied by heating of the stellar populations.}, and in view of
recent observations of non-zero mean stellar radial motions within the
disc. Indeed, using line-of-sight velocities of 213713 stars from the
RAVE survey, \cite{Siebert2011grad} found a Galactocentric radial
velocity gradient of $\lddp{V_R}{R} \simeq - 4\,{\rm km}\,{\rm
  s}^{-1}\,{\rm kpc}^{-1}$ in the extended Solar
neighbourhood. \cite{Siebert2012} found that such a gradient is
consistent with the effect of a $m=2$ quasi-static spiral density
wave, derived by \cite{LinShu}, although M16 showed that the reduction
factor is different in 3D. \cite{Monari2014} found the gradient
consistent with the effects of the Galactic bar, according to
test-particle simulations. Line-of-sight velocity fluctuations have
subsequently been detected on larger scales with red clump stars from
the APOGEE survey \citep{Bovy2015}, and we have then shown how the
peculiar velocity power spectrum of a $N$-body simulation with a
strong central bar and transient, co-rotating spiral arms fits very
well the observed power spectrum, while a quasi-stationary density
wave spiral model without a bar does not \citep{Grand2015}. Hereafter,
we will now check the peculiar velocity power spectrum in a simulation
coupling the effects of both a bar and a quasi-static spiral mode.

Not only non-zero mean radial motions have been found with recent
spectroscopic surveys, but also non-zero mean vertical motions
\citep{Widrow2012,Williams2013,Carlin2013}, which have amplitudes
$\lesssim 5\kmsec$ near the Galactic plane, but can reach $\sim
15\kmsec$ at large heights ($\sim 1.5\Kpc$), with a gradient of the
order of $\lddp{|V_z|}{z} \sim 10^{-2}\kmsec\pc^{-1}$ at the Solar
position. These typically consist in `breathing modes' or `bending
modes' of the disc. Breathing modes (bending modes) are vertical modes
with an odd (even) parity in the vertical velocity field and even
(odd) parity in the density distribution of the stars. It was shown
that any internal non-axisymmetric perturbations, such as the bar and
spiral arms, naturally cause breathing modes \citep[][hereafter F14,
  M15, and M16]{Faure2014,Monari2015,Monari2016}. Nevertheless, it
appears that the mean vertical motions induced by the bar in the Solar
vicinity are much smaller than the observed ones (M15). Those linked
to spiral arms are more important (F14, M16), due to a more rapid
radial variation of their potential, but still need unrealistically
large amplitudes to reproduce the observed mean motions. Hence, we
investigate here the effects of coupled bar and spirals in
test-particle simulations, to test whether the vertical motions arise
from a simple linear addition of the isolated effects of both, as it
is expected to be the case for in-plane motions.

In \Sec{sect:sim}, we describe the set-up of our test-particle
simulations with a bar, a fiducial spiral pattern (30 per cent density
contrast) and strong spiral pattern (60 per cent density contrast), and
simulations taking into account both the bar and spirals
simultaneously. We analyze the power spectra of peculiar line-of-sight
velocities in \Sec{sect:powlos}. We then analyze the detailed velocity
field of each simulation in \Sec{sect:res}, the effect of the coupling
and resonance overlaps on disc churning in \Sec{sect:radm}, and make
predictions for forthcoming spectroscopic surveys in
\Sec{sect:surveys}. We conclude in \Sec{sect:concl}.

\section{Set-up of simulations}\label{sect:sim}
In the following test-particle simulations we integrate forward in
time the equations of motion of massless particles (representing the
stars of the Milky Way disc) moving in a gravitational potential
(representing the potential of the Milky Way and its
non-axisymmetries), uninfluenced by the particles themselves.

\subsection{Potential}\label{sect:pot}
The potential that we use to represent the gravitational field of the
Milky Way is composed of an axysimmetric part and (i) of a bar, or
(ii) of a two-armed spiral pattern, or (iii) of both. In the
following, we use the Galactocentric cylindrical coordinates
$\pare{R,\phi,z}$, and velocities
$\pare{v_R,v_\phi,v_z}\equiv\pare{\dot{R},R\dot{\phi},\dot{z}}$.

The axisymmetric part of the potential corresponds to Model~I by
\cite{BT2008}, fitting several of the properties of the Milky Way
structure and consisting of two spheroidal components, a dark halo and
a bulge, and three disc components: thin, thick, and ISM disc. The
mass of the dark halo inside $100\Kpc$ is
$M_{\mathrm{h},<100\Kpc}=6\times 10^{11}\Msun$, and the total mass of
the bulge is $M_\mathrm{b}=5.18\times 10^{9}\Msun$. The disc densities
are exponential both in $R$ and in $z$. In particular the radial scale
length of the thin and thick disc is $\Rd=2\Kpc$, and their scale
heights $h_z^{\mathrm{thin}}=0.3\Kpc$ and
$h_z^{\mathrm{thick}}=1\Kpc$. The ISM disc has scale length and height
$\Rd^{\mathrm{ISM}}=4\Kpc$ and $h_z^{\mathrm{ISM}}=0.08\Kpc$
respectively, and a hole for $R<4\Kpc$. The total mass of the three
disc components is $M_{\mathrm{d}}=5.13\times 10^{10}\Msun$. In this
model the Sun is placed at $\pare{R_0,\phi_0,z_0}=\pare{8\Kpc,0,0}$,
i.e., we measure the angles from the line connecting the Sun and the
center of the Galaxy.

The bar potential is a 3D version of the pure quadrupole model used
by, e.g. \cite{Weinberg1994} and \cite{Dehnen2000}. It reads
\begin{equation}
  \potb(R,\phi,z,t)=\alpha\frac{v_0^2}{3}\pare{\frac{R_0}{\Rb}}^3
  U(r)\frac{R^2}{r^2} \cos\gammab,
\end{equation}
where $r^2=R^2+z^2$ is the spherical radius, $R_b$ is the length of
the bar, $R_0$ is the Galactocentric radius of the Sun, and $v_0$ is
the circular velocity at $R_0$,
\begin{equation}
  \gammab\pare{\phi,t}\equiv 2\left(\phi-\phib-\Omegab
    t\right),
\end{equation}
and
\begin{equation}
  U(r) \equiv \left\{
  \begin{array}{l l}
    (r/\Rb)^{-3} & \quad \text{for } r \geq \Rb,\\
    (r/\Rb)^{3}-2 & \quad \text{for } r < \Rb.
  \end{array} \right.
\end{equation}
The amplitude $\alpha$ is the ratio between the bar's and axisymmetric
contribution to the radial force, along the bar's long axis at
$(R,z)=(R_0,0)$. We choose for the simulations $\alpha=0.01$ as in
\cite{Dehnen2000}, $\Omegab=52.2\kmseckpc$ so that
$\Omegab/\Omega(R_0)=1.89$ \citep{Antoja2014}, $\Rb=3.5\Kpc$, and
$\phib$ such that at the end of the simulations $\te$, the bar major
axis has a $25\degr$ inclination w.r.t. the line connecting the Sun
and the center of the Galaxy, i.e., $\phib+\Omegab\te=25\degr$
(\citealt{Dehnen2000}), where $t_e$ corresponds to the present
time. For comparison, in one case (see below) we will present results
in the case where $\phib+\Omegab\te=45\degr$. The bar constructed in
this way does not modify the Galaxy's total mass and circular velocity
curve. The mass of the baryons going into the bar (equal to the
integrated positive density part of the bar) is about $4.43\times
10^9\Msun$.

The bar considered here is slightly different from the one of
\cite{Bovy2015} in terms of amplitude (our bar is 50 per cent weaker),
and vastly different from the one of \cite{Wegg2015} for structure
(half-length of the bar $5\Kpc$) and pattern speed \citep[$\lesssim
  45\kmsec$, see also][]{Portail2015}.

  It could be that the Galaxy's spiral pattern is composed of multiple
  modes with different pattern speeds
  \citep[e.g.][]{Quillen2011,SellwoodCarlberg2014}, but observations
  indicate that the non-axisymmetric part of the old stellar component
  of the Milky Way disc is dominated by a two-armed spiral pattern
  outside of the bar region \citep[namely, the Scutum-Centaurus and
  Perseus arms,][]{Benjamin2005,Churchwell2009}. Therefore, we
  consider a spiral perturbation consisting in a two-armed model, in
  the form originally proposed by \cite{CoxGomez2002}
\begin{equation}
  \Phis(R,\phi,z,t)=-\frac{A}{\Rs K
    D}\cos\gammas\left[\sech\left(\frac{Kz}{\beta}\right)\right]^\beta,
\end{equation}
where
\begin{subequations}
\begin{equation}
  K(R)=\frac{2}{R\sin p},
\end{equation}
\begin{equation}
  \beta(R)=K(R)\hs\paresq{1+0.4K(R)\hs},
\end{equation}
\begin{equation}
  D(R)=\frac{1+K(R)\hs+0.3\paresq{K(R)\hs}^2}{1+0.3K(R)\hs},
\end{equation}
\begin{equation}
  \gammas\pare{R,\phi,t}=2\left[\phi-\phis-\Omegas t+\frac{\ln(R/\Rs)}{\tan p}\right].
\end{equation}
Here, $p$ is the pitch angle, $A$ the amplitude of the spiral
potential, $h_s$ controls the scale-height of the spiral, and $R_s$ is
the reference radius for the angle of the spirals. This potential
corresponds to a spiral density distribution
\begin{equation}
  \rhos(R,\phi,z,t)\approx\rho_0
  \frac{K\hs}{D}\frac{\beta+1}{\beta}\cos\gammas\left[\sech\left(\frac{Kz}{\beta}\right)\right]^{2+\beta},
\end{equation}
\end{subequations}
where $\rho_0=A/(4\upi G \Rs \hs)$. We choose $\Rs=1\Kpc$,
$\Omegas=18.9\kmseckpc$, $\phis+\Omegas\te=-26\degr$, and
$p=-9.9\degr$ (\citealt{Siebert2012}; F14; M16). Moreover, we specify
two values for $A$: the first corresponds to a $30$ per cent density
contrast of the spiral arms w.r.t. the background disc surface density
at $R_0$ (`reference spirals', $A=341.8\kmsect$), the second to a
$60$ per cent density contrast (`strong spirals',
$A=683.7\kmsect$). With these values of $A$, the spiral arms produce a
maximum radial force of $0.5$ per cent (reference spirals) and $1$ per
cent (strong spirals) of the force due to the axisymmetric background
at $R=R_0$.

We label `B' the bar-only simulation, `S1' and `S2' the simulations
with only the reference and strong spirals respectively, and `BS1' and
`BS2' the simulations with the bar together with the reference and
strong spirals respectively. We will also show one case in which the
bar is coupled with the strong spirals and $\phib+\Omegab\te=45\degr$
(`B2S2'). The bar-only simulation, with the bar orientation
$\phib+\Omegab\te=45\degr$ is labelled `B2'
(Table~\ref{tab:sim_params}).
\begin{table}
  \begin{center}
  \begin{tabular}{llll}
    \hline
    Simulation & bar's $\alpha$ & spiral's $A\pare{\kmsect}$ & $\phib+\Omegab\te$ \\ 
    \hline
    B & $0.01$ & $0$ & $25\degr$ \\
    B2 & $0.01$ & $0$ & $45\degr$ \\
    S1 & $0$ & $341.847$ & - \\
    S2 & $0$ & $683.694$ & - \\
    BS1 & $0.01$ & $341.847$ & $25\degr$ \\
    BS2 & $0.01$ & $683.694$ & $25\degr$\\
    B2S2 & $0.01$ & $683.694$ & $45\degr$ \\
    \hline
  \end{tabular}
  \caption{Amplitudes of the perturbation potentials presented in this
    work.}
  \label{tab:sim_params}
  \end{center}
\end{table}

\begin{table}
  \begin{center}
  \begin{tabular}{ll}
    \hline
    Resonance & $R$(kpc) \\ 
    \hline
    Bar's corotation & 4.08 \\
    Bar's OLR  & 7.22  \\
    Spiral arms' corotation & 11.49 \\
    Spiral arms' ILR & 1.89  \\
    Spiral arms' inner 4:1 resonance & 7.61  \\
    Spiral arms' inner 3:1 resonance & 6.09  \\
    \hline
  \end{tabular}
  \caption{Radii of the resonances in all considered models in this work.}
  \label{tab:resonances}
  \end{center}
\end{table}

\subsection{Initial conditions}
We generate the disc initial conditions as discrete realizations of
the phase-space Shu-Schwarzschild distribution function
(\citealt{Shu1969,BienaymeSechaud1997,BT2008}), as explained in
F14. In this way we obtain a disc of $5\times10^7$ particles whose
surface density distribution in configuration space is approximately
exponential with radius (scale length $\Rd$), and whose vertical
distribution is determined by the restoring force. The radial and
vertical velocity dispersion on the disc plane vary approximately as
$\sigma_R\approx\sigma_{R,0}\exp\paresq{-\pare{R-R_0}/\Rsg}$,
$\sigma_z\approx\sigma_{z,0}\exp\paresq{-\pare{R-R_0}/\Rsg}$, where
$(\sigma_{R,0},\sigma_{z,0})=(35,15)\kmsec$, and $\Rsg=5\Rd$
(\citealt{BienaymeSechaud1997}, F14). The test particles distribution
obtained with these parameters resemble the thin disc of
\Sec{sect:pot}, which is the only stellar disc modeled with test
particles.

\subsection{Time scales}
We integrate forward our initial conditions for a total time
$T=9\Gyr$, from $\ti=-3\Gyr$ to $\te=6\Gyr$. 

\begin{figure}
  \centering 
  \includegraphics[width=\columnwidth]{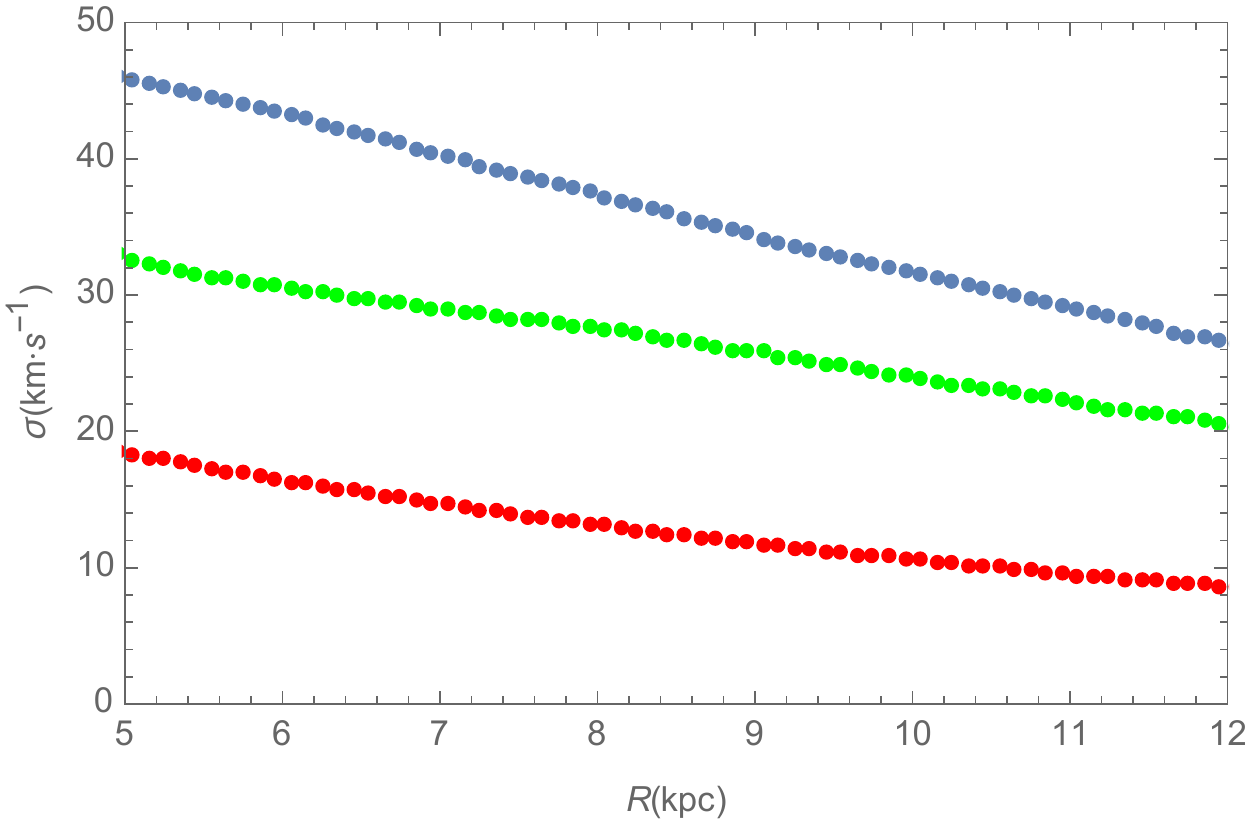}
  \caption{Velocity dispersion of the disc test particles with
    $|z|<0.5\Kpc$ at $t=0$, as a function of $R$. Blue dots: radial
    velocity dispersion $\sigma_R$. Green dots: tangential velocity
    dispersion $\sigma_\phi$. Red dots: vertical velocity dispersion
    $\sigma_z$.}
  \label{fig:disp}
\end{figure}

\begin{figure}
  \centering 
  \includegraphics[width=\columnwidth]{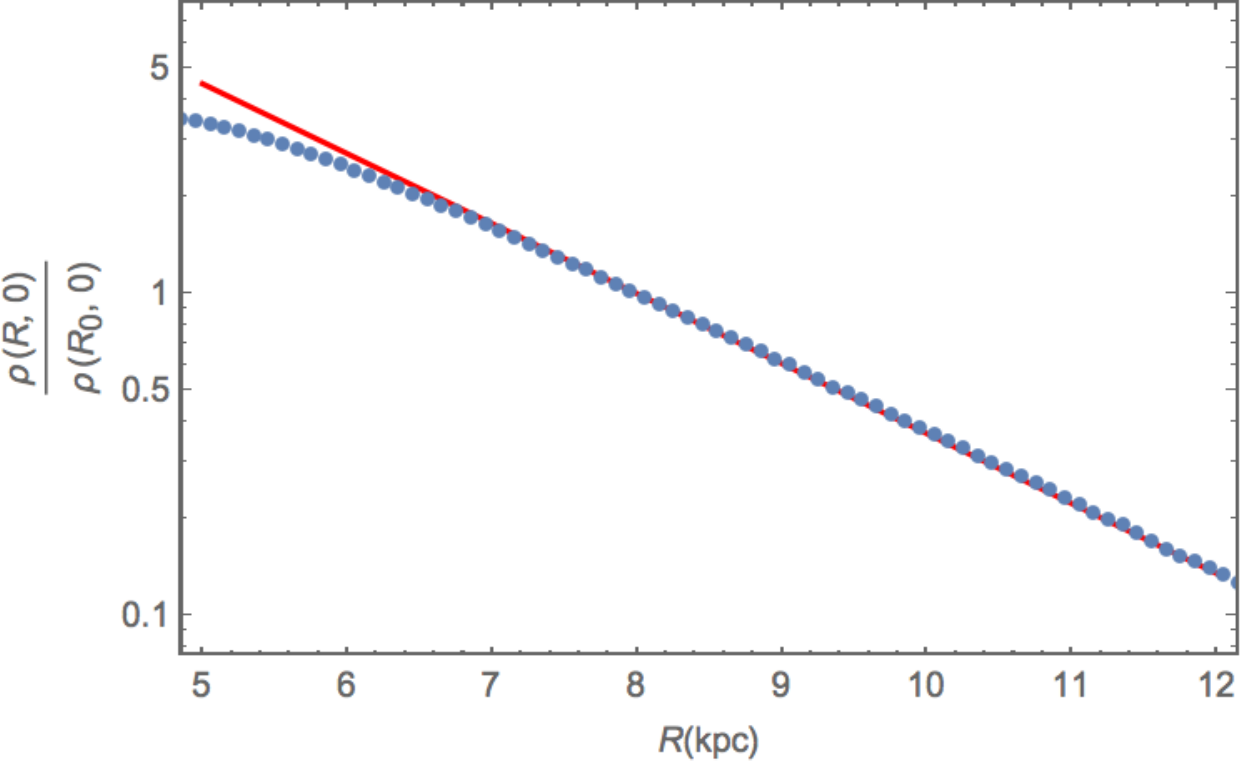}
  \includegraphics[width=\columnwidth]{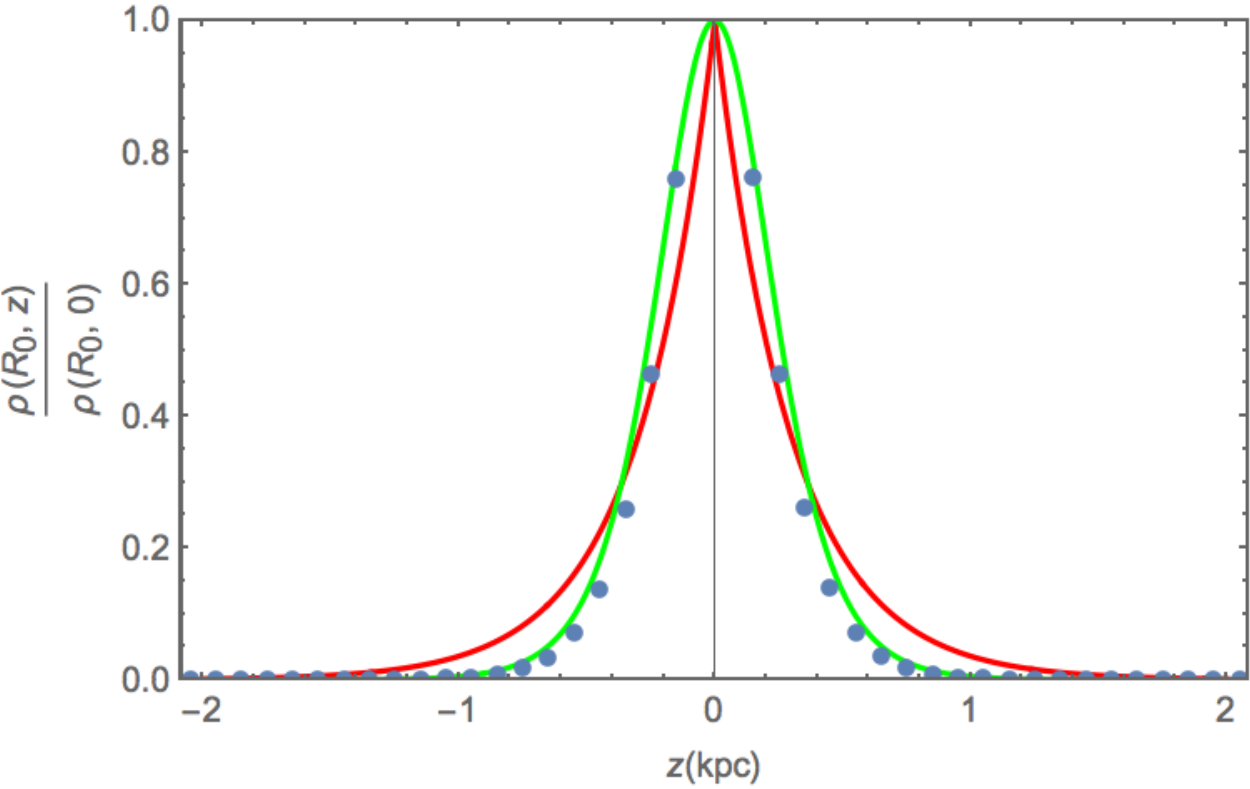}
  \caption{Profiles for the disc test particles density at $t=0$
      (blue dots), and the volume density $\rho_{\mathrm{thin}}$ of
      the thin disc (red line) of Model~I by \protect\cite{BT2008}, used here as
      a background potential. Top panel: radial density
      profile of particles with $|z|<0.3\Kpc$ compared with
      $\rho_{\mathrm{thin}}(R,0)$. Bottom panel: vertical density
      profile of particles with $|R-R_0|<0.3\Kpc$ (blue dots) compared with
      $\rho_{\mathrm{thin}}(R_0,z)$ (red line); the green line is the function
      $f(z)=\sech^2(z/(0.3\Kpc))$.}
  \label{fig:den}
\end{figure}

For $t<0$ we integrate the initial conditions in the axisymmetric part
of the potential only. We do this so that the initial conditions
become mixed with the background potential\footnote{Since the
  Shu-Schwarzschild distribution function is built on approximated
  integrals of motion, because of Jeans' Theorem, it will evolve. The
  faster, the worse the approximation of the integrals.}. After the
initial $3\Gyr$, we obtain stable distribution functions in the
Model~1 of \cite{BT2008} potential, with velocity dispersions at
$(R,z)=(R_0,0)$:
$(\sigma_R,\sigma_\phi,\sigma_z)\approx(37,27,13)\kmsec$ (see
\Fig{fig:disp}). The vertical restoring force determines the $z$
density profile. At $R=R_0$ this is nicely fitted by a $\sech^2$
profile with scale height $h_z\approx0.3\Kpc$. In \Fig{fig:den} we
compare the density of test particles after the mixing time with the
density of the thin disc in the background potential. The radial
profiles (top panel) are in excellent agreement over a large range of
radii, and diverge only in the central parts because of the tapering
of the initial conditions at $R<4\Kpc$ (see F14). The vertical profile
(bottom panel) differs only slightly from the one of the thin disc of
the background potential which is purely exponential, because of the
form of the distribution function used to generate the test particles
initial conditions.

The bar and the spiral arms forces are present in the simulations only
for $t>0$. These are introduced in the simulations by letting their
respective amplitudes $\alpha$ and $A$ grow by a factor
(\citealt{Dehnen2000})
\begin{equation}
  \eta(t)=\left(\frac{3}{16}\xi^5-\frac{5}{8}\xi^3+\frac{15}{16}\xi+\frac{1}{2}\right),\quad
  \xi\equiv2\frac{t}{3\Gyr}-1,
\end{equation}
until, at $t=3\Gyr$ they reach their final amplitude which is kept
constant until the end of the simulation at $t=\te\equiv 6\Gyr$.

\section{Peculiar l.o.s. velocity power spectrum}\label{sect:powlos}
In a recent paper, \cite{Bovy2015} proposed a method to characterise
the peculiar velocity field of the disc stars
% constrain some of the fundamental parameters of the Milky Way 
using a pencil beam survey, where the spectroscopic information
(i.e. the line of sight $\vlos$ velocity) and the photometric distance
of the stars are estimated \citep[as it happens for the APOGEE
survey,][]{APOGEE}. This method analyses
 %relies on the minimization of the
the power spectrum of the peculiar line of sight (l.o.s.) velocity
(see below for a definition), and the observed power spectrum
was tested by \cite{Bovy2015} on some simple simulations, including a
bar or spiral arms. These tests showed a striking difference in the
spectra induced by the bar and spiral arms models considered.

\cite{Grand2015} computed the power spectrum of the kinematics of more
sophisticated $N$-body models of disc galaxies and found that this is
sensitive to parameters such as the number of spiral arms, spiral arm
pitch angle, and position of the Sun with respect to the spiral
arm. In particular, they consider the power spectrum of the peculiar
line of sight velocity projected on the Galactic plane, defined as
\begin{eqnarray}\label{eq:vlos}
  \vlost \elos &\equiv& \avvR \cos b \cos(\upi-\phi-l)\eR + \nonumber
  \\ && (\avvphi-\avvphiz)\cos b \sin(\upi-\phi-l)\ephi,
\end{eqnarray}
where 
\begin{equation}\label{eq:avvphi}
  \avvphiz\equiv\frac{1}{2\upi}\int_0^{2\upi}\avvphi~\de\phi,
\end{equation}
$\eR$, $\ephi$, and $\elos$ are the versors in the radial,
tangential and l.o.s. direction, $l$ and $b$ are the Galactic
longitude and latitude, and the averages $\av{\cdot}$ are computed at
some position $(R,\phi)$ of the Galactic plane. In this work $l$ and
$b$ are computed assuming the Sun at $(R,\phi,z)=(8\Kpc,0,0)$. The
quantity $\vlost$ represents the difference between the mean
l.o.s. velocity and the mean l.o.s. velocity of the axisymmetric
background. To compute the power spectrum, $\vlost$ is evaluated on a
$N\times N$ grid in $x\equiv R\cos\phi$ and $y\equiv R\sin\phi$, so
that the power $P(k_x,k_y)$ is
\begin{equation}
  P\pare{k_x,k_y}=\pare{4\upi}^2\absp{F\pare{k_x,k_y}}^2,
\end{equation}
where $F\pare{k_x,k_y}$ is the discrete Fourier transform of
$\vlost(x,y)$ on the $N\times N$ grid ($k_x$ and $k_y$ are in units of
$\Kpc^{-1}$, and $P$ in units of $\Km^2\mathrm{s}^{-2}$). In
particular, $P\pare{k_x,k_y}$ is estimated at the wavenumbers
$\pare{k_x,k_y}=(\kmax l/N,\kmax m/N)$, with $l=1,...,N$, $m=1,...,N$,
and $\kmax$ is the Nyquist wavenumber. The one-dimensional power
spectrum $P(k)$ is computed by averaging $P\pare{k_x,k_y}$ along
rings in the $k_x-k_y$ plane \citep[for the details
  see][]{Bovy2015,Grand2015}.

  Here, as in \cite{Bovy2015} and \cite{Grand2015} we estimate $P(k)$
  in a portion of the disc resembling the one spanned from the APOGEE
  survey, i.e., $x \in \paresq{5.5\Kpc,12.5\Kpc}$, $y \in
  \paresq{-3.5\Kpc,4.5\Kpc}$, and $\absp{z}<0.25\Kpc$. The $x-y$ grid
  has bin size $0.8\Kpc$.
\begin{figure}
  \centering \includegraphics[width=\columnwidth]{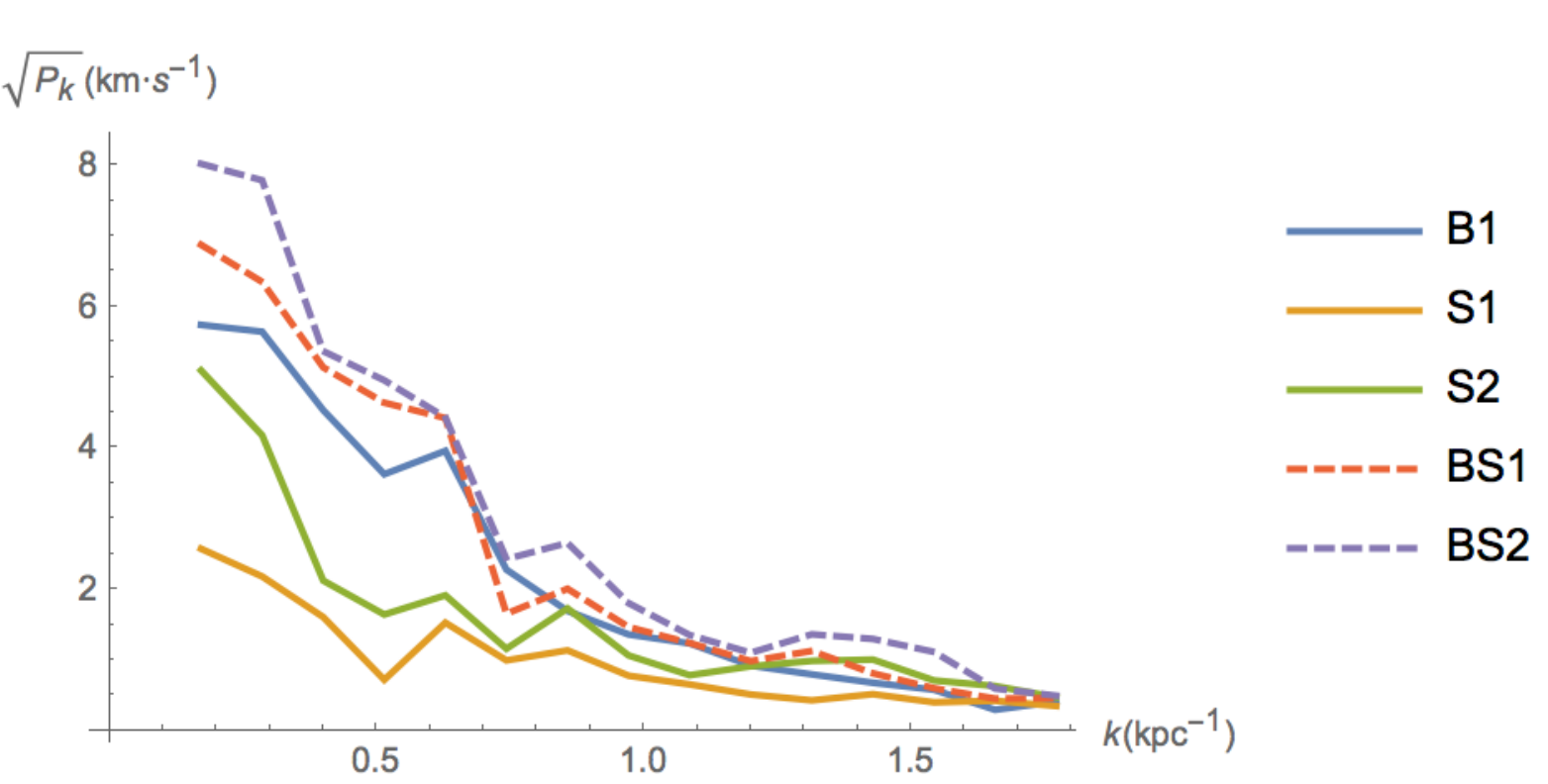}
  \includegraphics[width=\columnwidth]{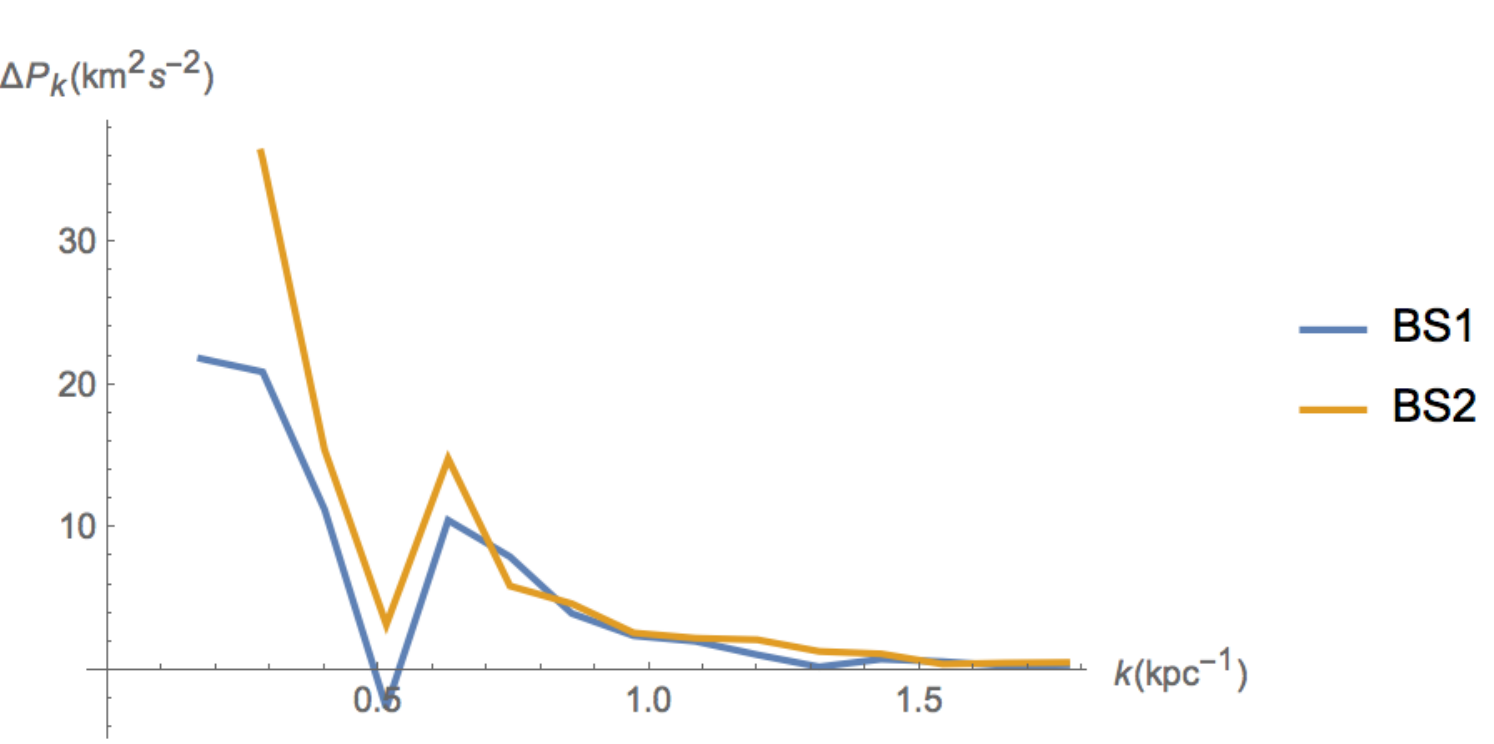}
  \caption{Top: power spectrum $\sqrt{P_k}$ for all the simulations,
    computed in a volume similar to the one spanned by APOGEE. Bottom:
    $\Delta P_k$ for the BS1 and BS2 cases respectively.}
  \label{fig:Pk}
\end{figure}
In \Fig{fig:Pk} (top) we show this power spectrum for the several
models. The first thing we note is that while the spiral models (S1
and S2) bear some resemblance with \cite{Bovy2015} spiral models (even
though they do not peak as strongly at $k \lesssim 1\Kpc^{-1}$), their
bar model has a very strong peak at $k\sim 0.5\Kpc^{-1}$, that is not
present in the spectra of this work\footnote{The same peak is also
  present in the barred model presented in \cite{Grand2015}.}. We
associate these differences both to the fact that their fiducial bar
model grows faster (their slowly grown model is more similar to ours),
and that their perturbation is $50$ per cent times stronger than ours
\citep[and of][]{Dehnen2000}. In fact, all our spectra resemble more
their $m=2$ elliptical perturbations, with most of the power on large
scales. Linear theory predicts that, at least away from the resonances
(Table~\ref{tab:resonances}), $\vlost$ scales linearly with the
strength of the perturbation, and so do
$\sqrt{P_k}\propto|\vlost|$. Therefore, the strength of the
perturbation mostly influences the amplitude of the power
spectrum. Another significant factor in shaping the power spectrum,
and in particular the relative height of the peaks, is the volume of
configuration space on which the power spectrum is computed. This
might explain differences with the APOGEE data points. \cite{Bovy2015}
compute the power spectrum of their synthetic models with the bar in
the same volume of configuration space as the one we use, but the time
scales of their simulations are different. This can cause significant
non-stationary features in the velocity field (see Section~4.1) that
disappear as the disc gets more phase-mixed, and which can influence
the power spectrum. Notice that, while the circular velocity curve and
position of the resonances in the synthetic models of \cite{Bovy2015}
are similar to those in the simulations of the present work, other
differences could contribute to create a different power spectrum: the
2D nature of the simulations of \cite{Bovy2015} versus the 3D nature
of our simulations, a slightly kinematically colder disc
($\sigma_R(R_0)=31.4\kmsec$ in their case), and a longer scale length
($\Rd=3\Kpc$). Using the linear theory of M16 we found that a
$\Rd=3\Kpc$ scale length decreases the power on the largest scales,
while the differences with the $\sigma_R(R_0)=31.4\kmsec$ case are not
significant. However, these effects are small, if compared to the
differences due to a different volume for the computation of the power
spectrum, notably when comparing to APOGEE data and the intrinsic
selection function of the survey.

The models with the combination of bar and spiral arms present a
larger power of the peculiar l.o.s. velocity at all scales (but mostly
on large scales), with the model BS2 having larger power than
BS1. However, apart from the larger overall power of these models,
their spectra do not present striking qualitative differences with
respect to the other models (e.g. peaks of power at some particular
scale not present in the other cases). Therefore, we can ask ourselves
whether the power in the combined bar and spiral arms case is simply
the power obtained by linearly summing the single $\vlost$ fields
induced independently by the bar and spiral arms. Since the power goes
as the square of the peculiar velocity, we have
\begin{equation}\label{eq:cond}
  \Pk{BS}\leq \Pk{B}+\Pk{S}+2\sqrt{\Pk{B}\Pk{S}},
\end{equation}
where $\Pk{B}$, $\Pk{S}$, and $\Pk{BS}$ are the powers in the bar,
spiral arms, and coupled case respectively. In \Fig{fig:Pk} (bottom)
we plot the quantity
\begin{equation}
  \Delta P_k\equiv \Pk{B}+\Pk{S}+2\sqrt{\Pk{B}\Pk{S}}-\Pk{BS}.
\end{equation}
If $\Delta P_k\geq 0$ the condition \Eq{eq:cond} is respected. This
condition is \emph{necessary but not sufficient} to say that the
peculiar velocity is simply the sum of the peculiar velocity induced
by the bar and the spiral arms everywhere in the disc. \Fig{fig:Pk}
(bottom) shows that this condition is respected almost everywhere in
our simulations. We will hereafter turn to the detailed 3D velocity
field to see whether the superposition really is linear. Nevertheless,
the power spectrum in the fashion of \cite{Bovy2015} clearly indicates
that the APOGEE large-scale velocity fluctuations are probably
predominantly driven by the bar unless the spiral arm with a large
amplitude is a transient and corotating as suggested by
\cite{Grand2015}, and that the addition of spirals does not change
much, except for the addition of power on the largest scales. This
means that adding to a bar-only model quasi-static spirals with
density contrast as large as 60 per cent also provides an acceptable
fit. The amplitude of spirals certainly does not have to be negligible
for the power spectrum to match APOGEE observations. Also, this does
not prevent the spirals from having non-negligible influences at
specific radii linked to resonance overlaps (notably in terms of
churning), and on mean vertical motions, as we investigate hereafter
through a detailed analysis of the 3D velocity field.

\section{Analysis of the 3D velocity field}\label{sect:res}

\subsection{Bar simulation}
\begin{figure*}
  \centering
  \includegraphics[width=0.32\textwidth]{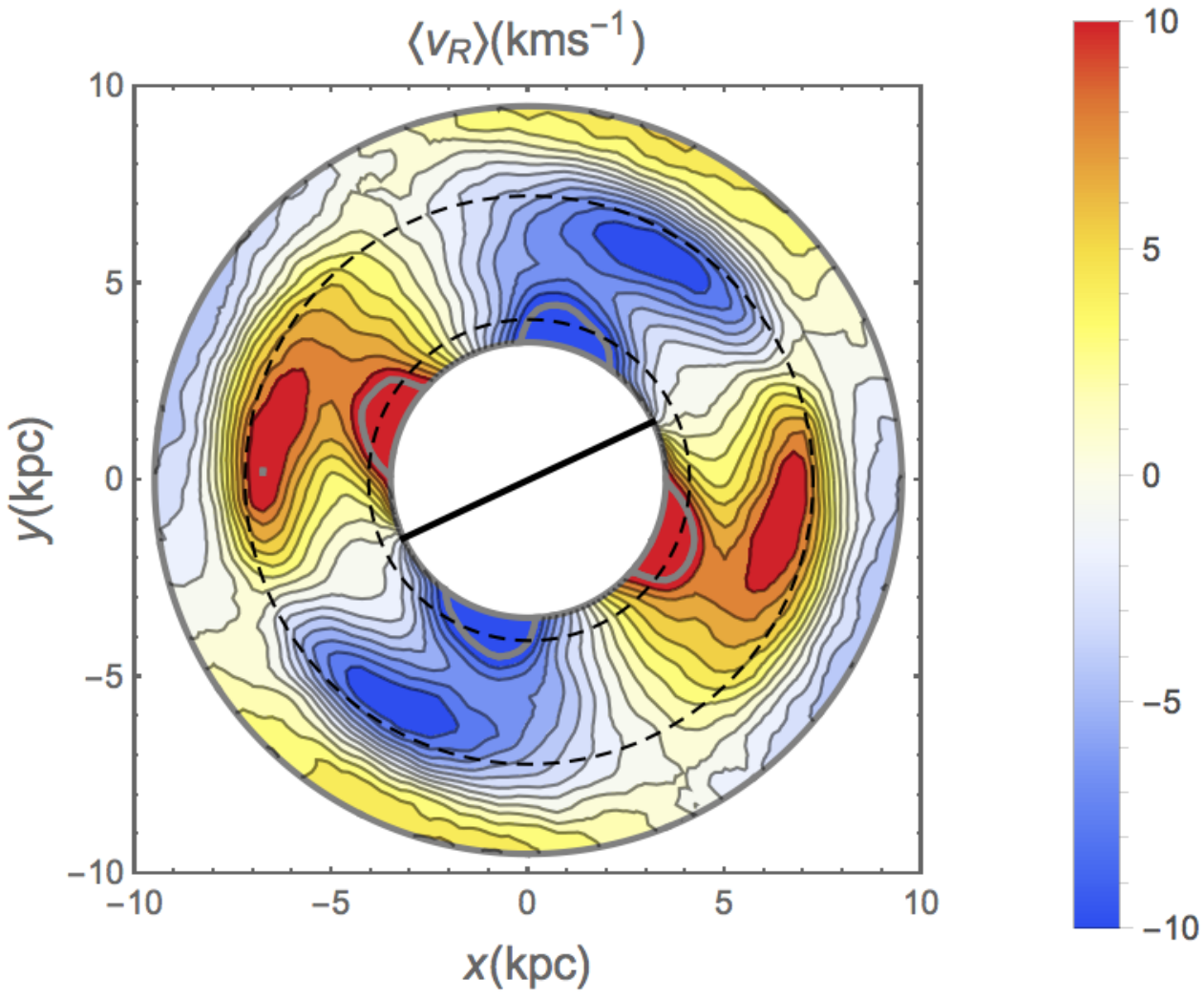}
  \includegraphics[width=0.32\textwidth]{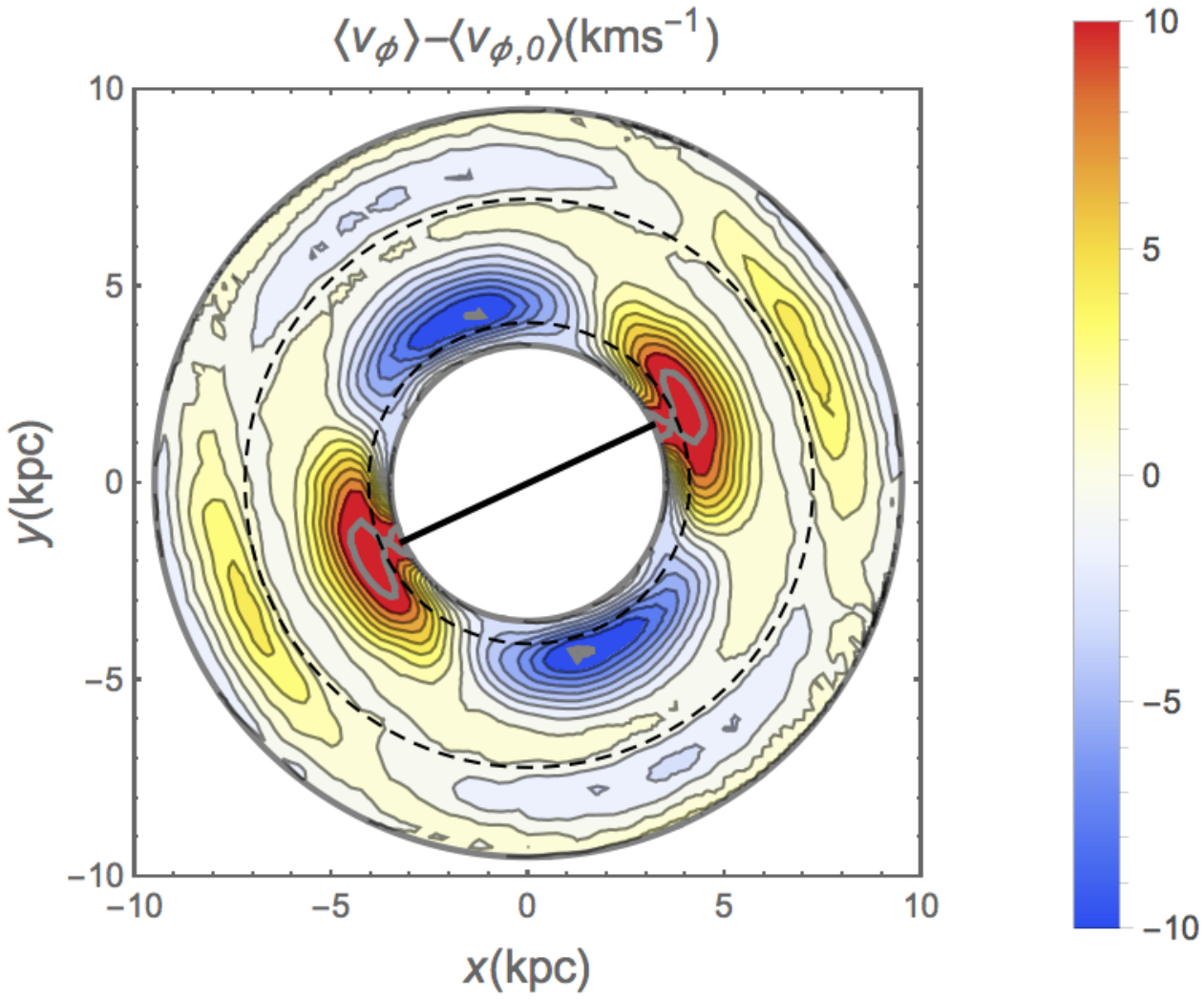}
  \includegraphics[width=0.32\textwidth]{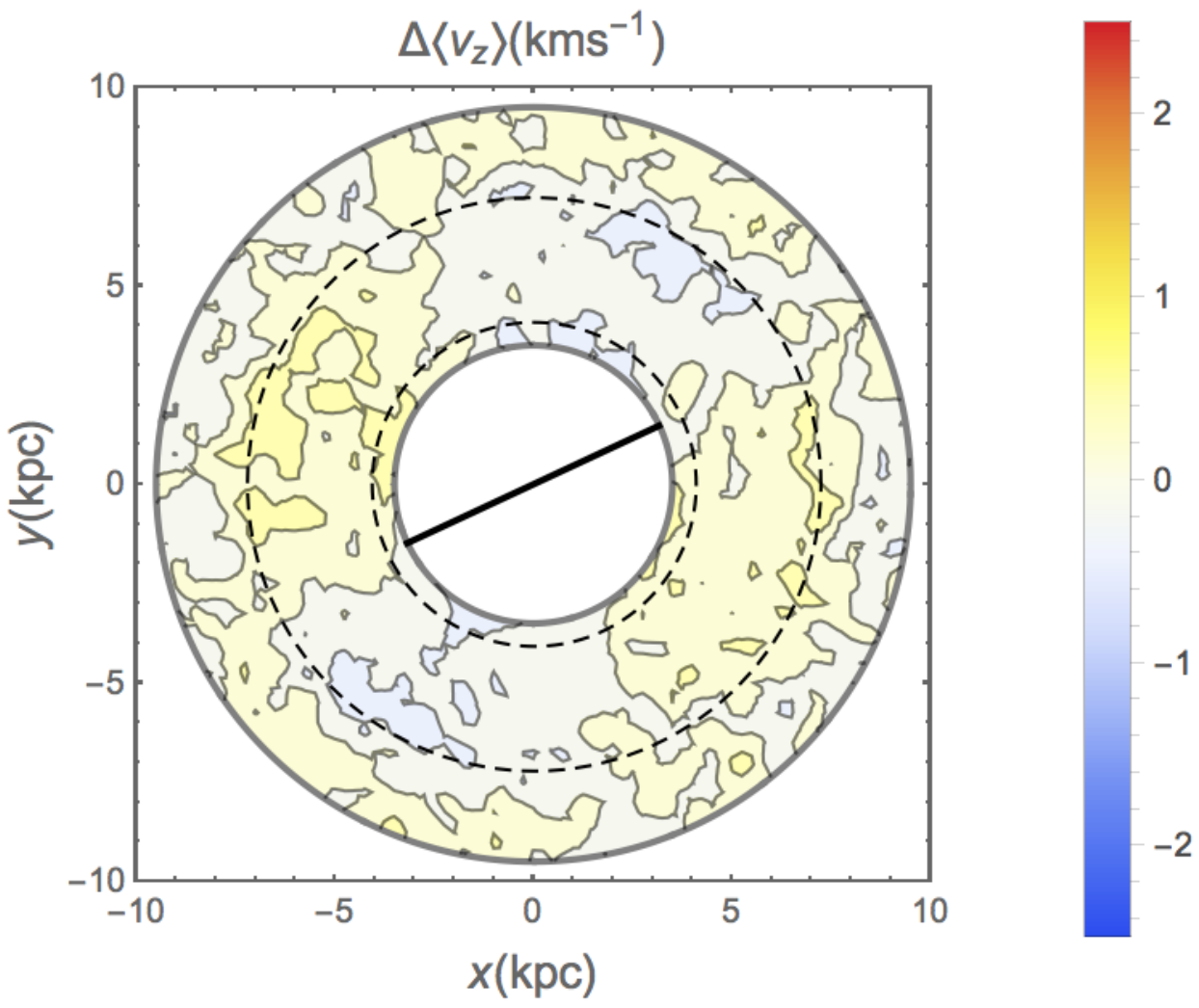}
  \caption{Average velocities at $t=\te$ for the simulation B on
    $x$~vs.~$y$ plane. Left panel: $\avvR$. Center: $\avvphi$. Right:
    $\dvz$. The averages are computed inside square bins of size
    $0.25\Kpc$. A Gaussian filter on a scale $0.5\Kpc$ is applied to
    the maps. The thick line at the center of the panels represents
    the long axis of the bar. The dashed circles represent the
    position of the corotation and OLR . The
    Galaxy rotates anti-clockwise.}
  \label{fig:bar}
\end{figure*}
In \Fig{fig:bar} we show the result of the simulation B. The 3 panels
represent the average of the velocity components of the particles at
$t=\te$ on the $\pare{x,y}\equiv\pare{R\cos\phi,R\sin\phi}$ plane,
inside square bins of size $250\pc$. The left panel represents
$\avvR$, the central panel $\avvphi-\avvphiz$, and the right the
quantity $\dvz$, i.e., the difference between $\avvz$ for $z>0$ and
for $z<0$ (see M15).

The average of the 3 velocity components has the form, at every $R$,
of a $m=2$ Fourier mode. These results are, at least far from the
resonances, in agreement with the findings of \cite{Kuijken1991} that
studied the response of the horizontal kinematics ($v_R$ and $v_\phi$)
to perturbations in 2D stellar discs, and with M15 that related the
mean $v_z$ to the average $v_R$ and $v_\phi$ in 3D, and noticed for
the first time that the bar induces non-zero mean vertical motions in
the whole Galaxy (even though of very small amplitude). At the
resonances (dashed circles) the effects of the perturbation are always
special. For $v_R$ we notice that the streaming motions induced by the
bar are particularly strong near the resonances, and that the phase of
the Fourier mode changes by $90\degr$ passing from inside to outside
the OLR . These effects can be related to
particular orbital configuration induced by the bar nearby the
resonances (M15, \citealt{BT2008}). The strongest effects on the
$v_\phi$ component are just outside the corotation. Notice how, just
outside the OLR, the behaviour of $v_\phi$ is not described by the
linear approximation, that would predict the maxima (minima) of
$\avvphi-\avvphiz$ aligned with the long (short) axis of the bar.
%a $m=2$ pure cosine mode with negative coefficient, while in this case
%it is positive for $~1-2\Kpc$, 
This prediction is eventually confirmed only at $R\sim9\Kpc$. This
behavior was also observed by \cite{Muhlbauer2003} in the case of
their hotter models. Finally, we plot the difference $\dvz$ between
the mean motion for particles with $z>0$ and $z<0$, as in M15, a
quantity which is positive (negative), i.e. corresponds to a
rarefaction (compression) for particles that move away from (towards)
the Galactic plane. In the case of B this breathing mode is always
quite moderate, and changes phase (of an angle of $90\degr$) at the
OLR, as predicted in M15.

\subsection{Spiral arms simulations}
\begin{figure*}
  \centering
  \includegraphics[width=0.32\textwidth]{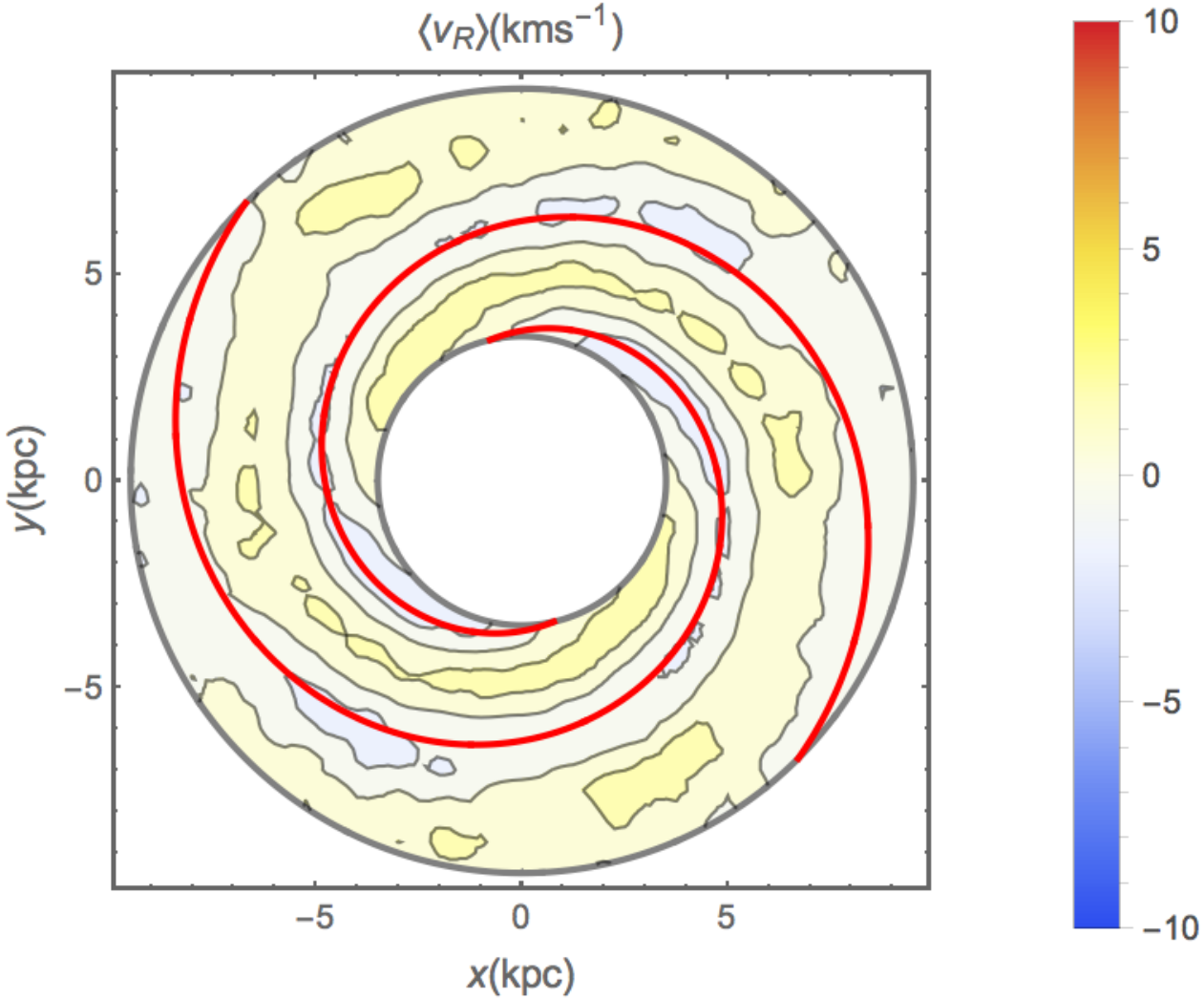}
  \includegraphics[width=0.32\textwidth]{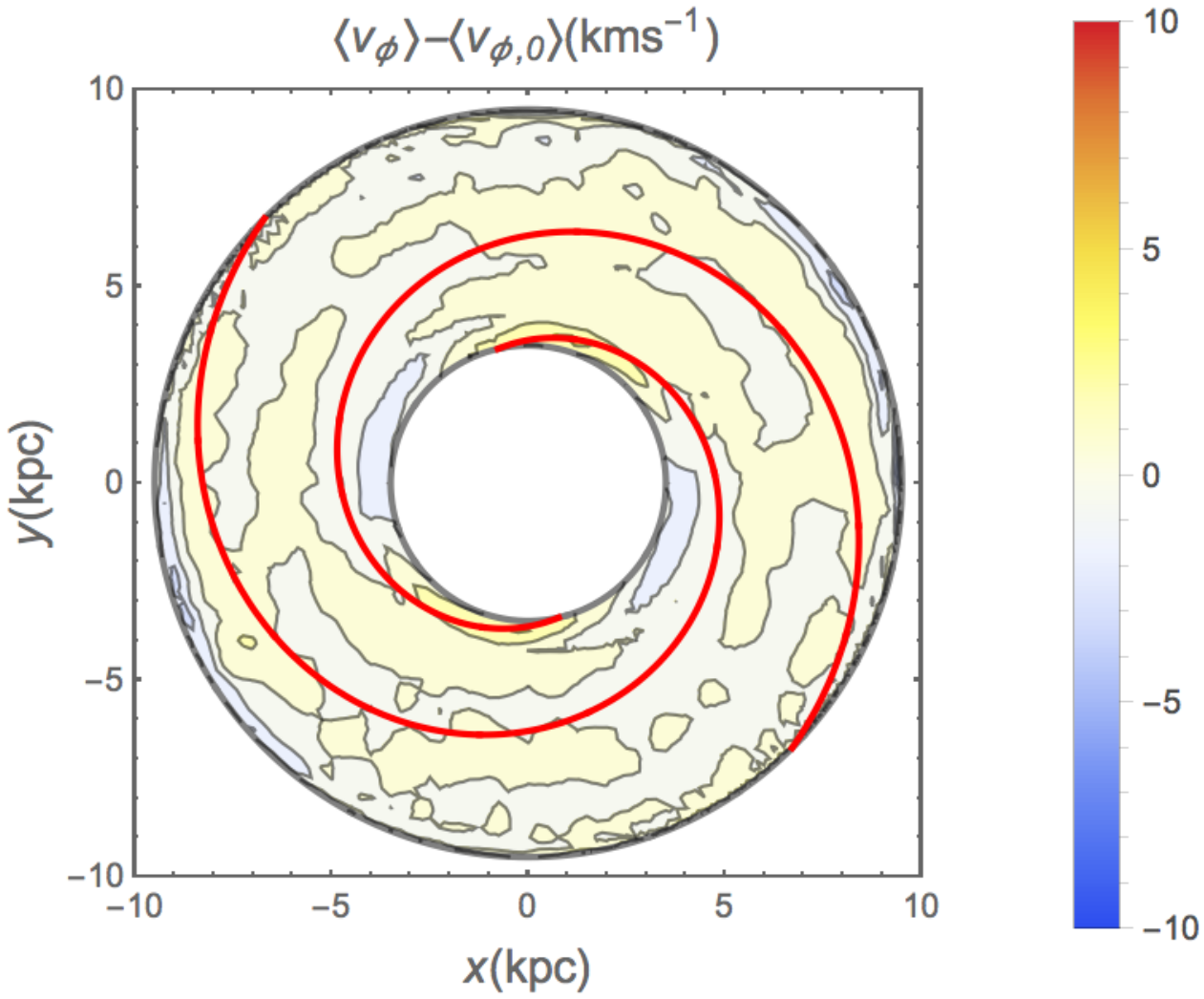}
  \includegraphics[width=0.32\textwidth]{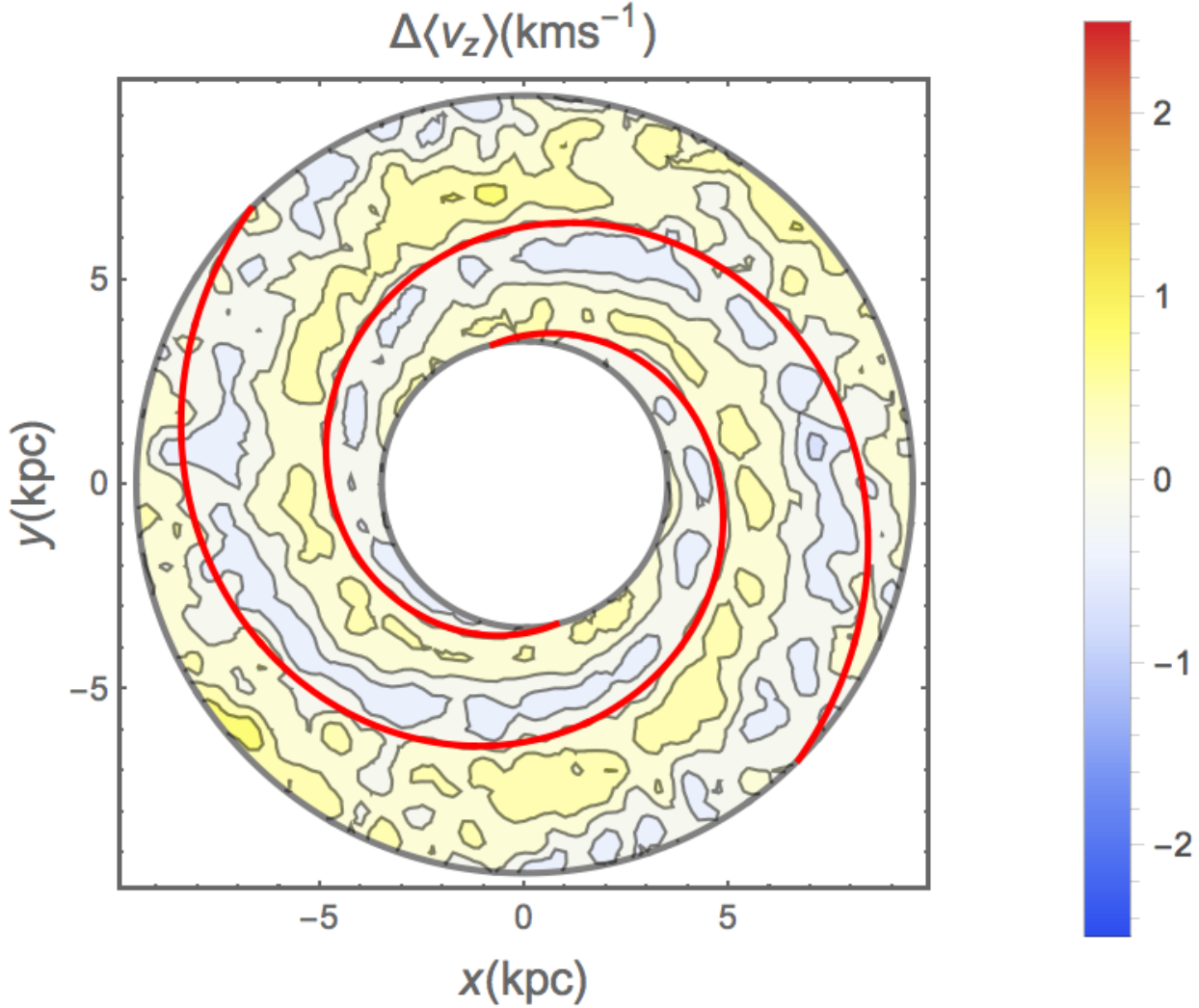}
  \caption{Average velocities at $t=\te$ for the simulation S1 on
    $x$~vs.~$y$ plane. Left panel: $\avvR$. Center: $\avvphi$. Right:
    $\dvz$. The averages are computed inside square bins of size
    $0.25\Kpc$. A Gaussian filter on a scale $0.5\Kpc$ is applied to
    the maps. The thick red line represents the locus of the spiral
    arms.}
  \label{fig:sp}
\end{figure*}
\begin{figure*}
  \centering
  \includegraphics[width=0.32\textwidth]{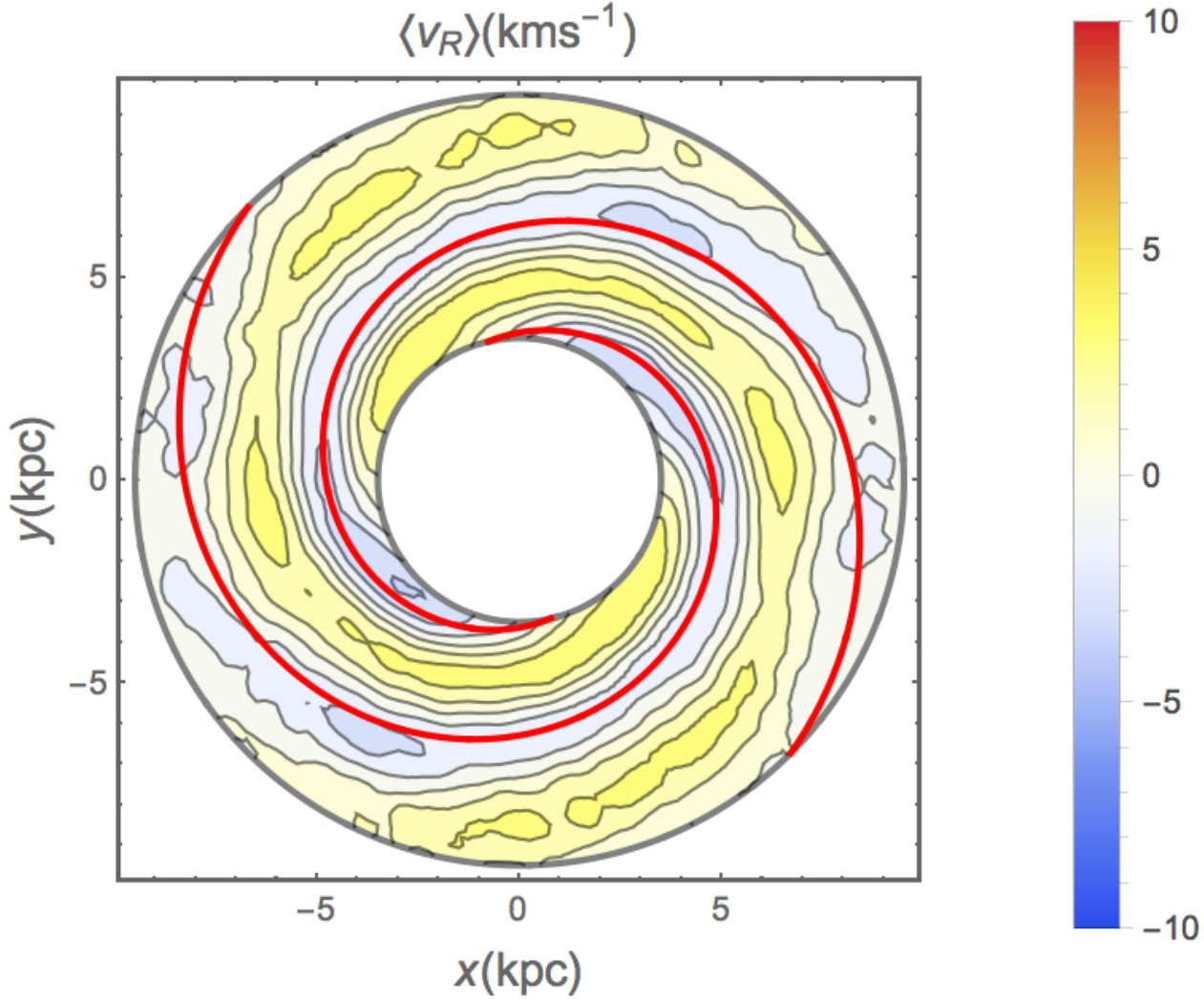}
  \includegraphics[width=0.32\textwidth]{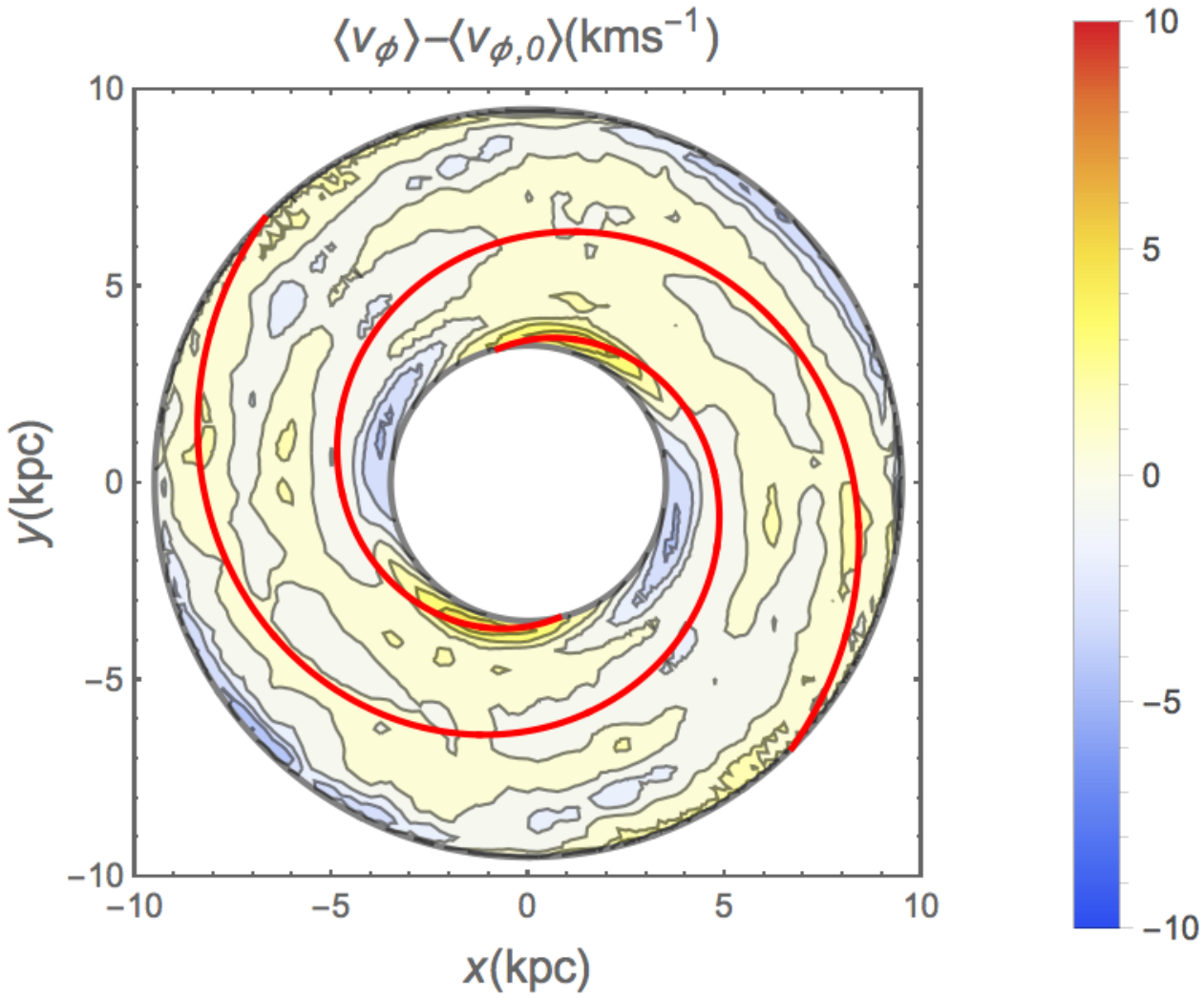}
  \includegraphics[width=0.32\textwidth]{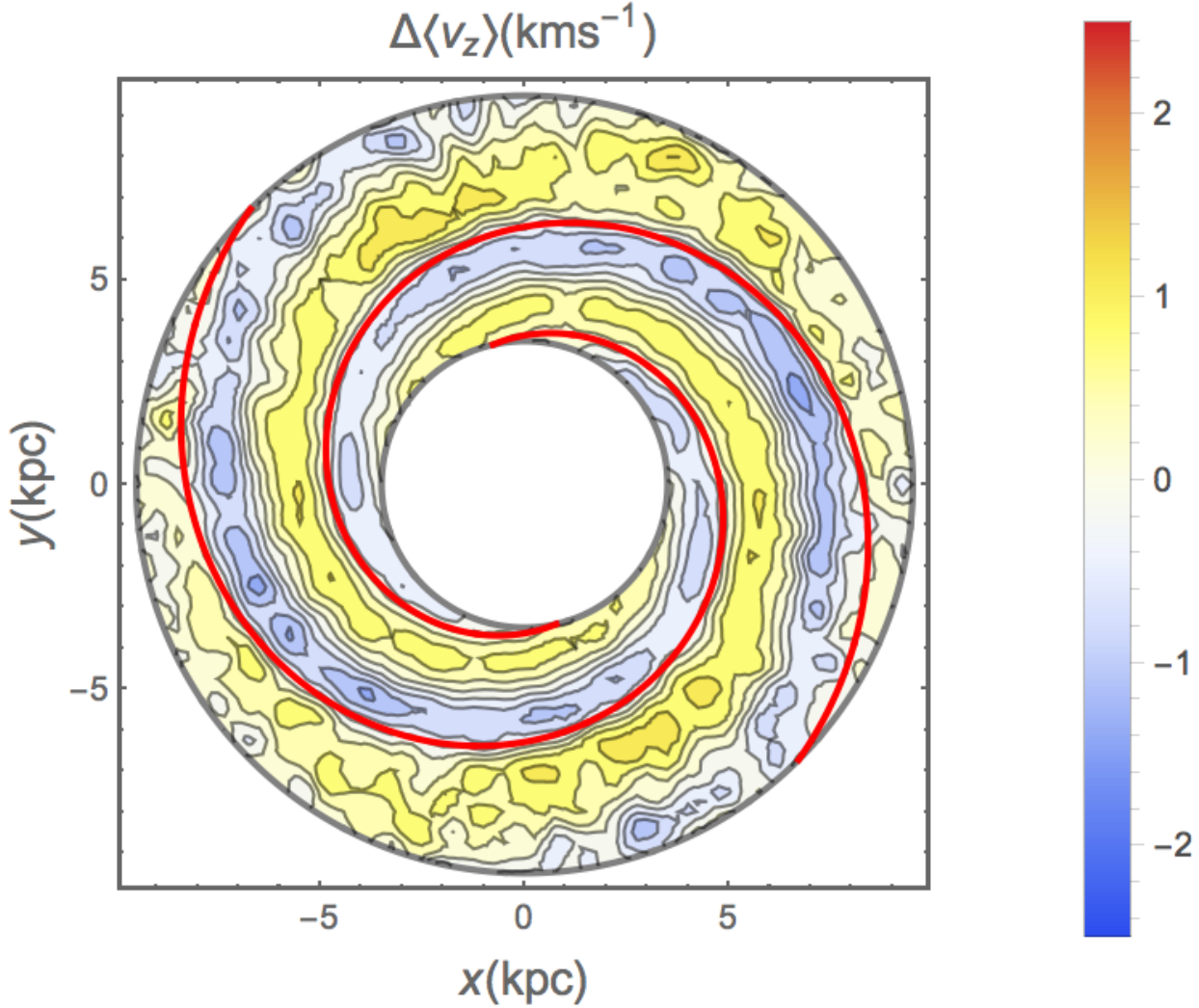}
  \caption{As in \Fig{fig:sp}, but for S2.}
  \label{fig:sp2}
\end{figure*}
The results of the S1 and S2 simulations show an agreement with the
models of M16 and a qualitative agreement with the simulations of F14
and \cite{Debattista2014}. \Figs{fig:sp}{fig:sp2} illustrate it, with
the same meaning of the left, central, and right panel as in
\Fig{fig:bar}. In these plots we only show the locus of the spiral
arms (thick red curves) since there are no major resonances in the
plotted regions. As already observed in F14 and M16, the average
velocity contours have the shape of two-armed Fourier modes resembling
to spiral arms. In particular, the locus of the spiral arms correspond
with the maximum amplitude of the $v_R$ streaming motion, directed
towards the center of the Galaxy in the arms region, and outwards in
the interarm regions (within corotation). The breathing mode is
$\dvz=0$ on the arms, and $\dvz<0$ ($\dvz>0$) in the trailing
(leading) edge of the arms (within corotation).  Contrary to the case
of the bar, there is thus a phase shift between the maxima of vertical
and radial bulk motions, a phase shift which is linked to the
oscillatory nature of the radial part of the spiral potential,
contrary to the case of the bar (see M15). The $v_\phi$ velocity (not
treated by F14) exhibits more complicated patterns, however the
general effect is a weak tendency for the stars to move faster outside
of spiral arms and slower inside.

The differences with F14 are quantitative. The kinematic responses are
much larger in their case. This is not surprising, considering that
the maximum radial force of the F14 spirals at $R_0$ and relative to
the axisymmetric background was $0.23$, while it is $0.05$ and $0.1$
for S1 and S2 respectively in the present case. Moreover, the vertical
force exerted by the F14 spiral arms is much stronger than in the
present cases. Note that the \cite{CoxGomez2002} model that we use is
more realistic, as it is the one related by the Poisson equation to
spiral arms that have a realistic density $\approx\sech^2$ density
fall-off in $z$.

\subsection{Coupled bar-spiral simulations}
\begin{figure*}
  \centering
  \includegraphics[width=0.32\textwidth]{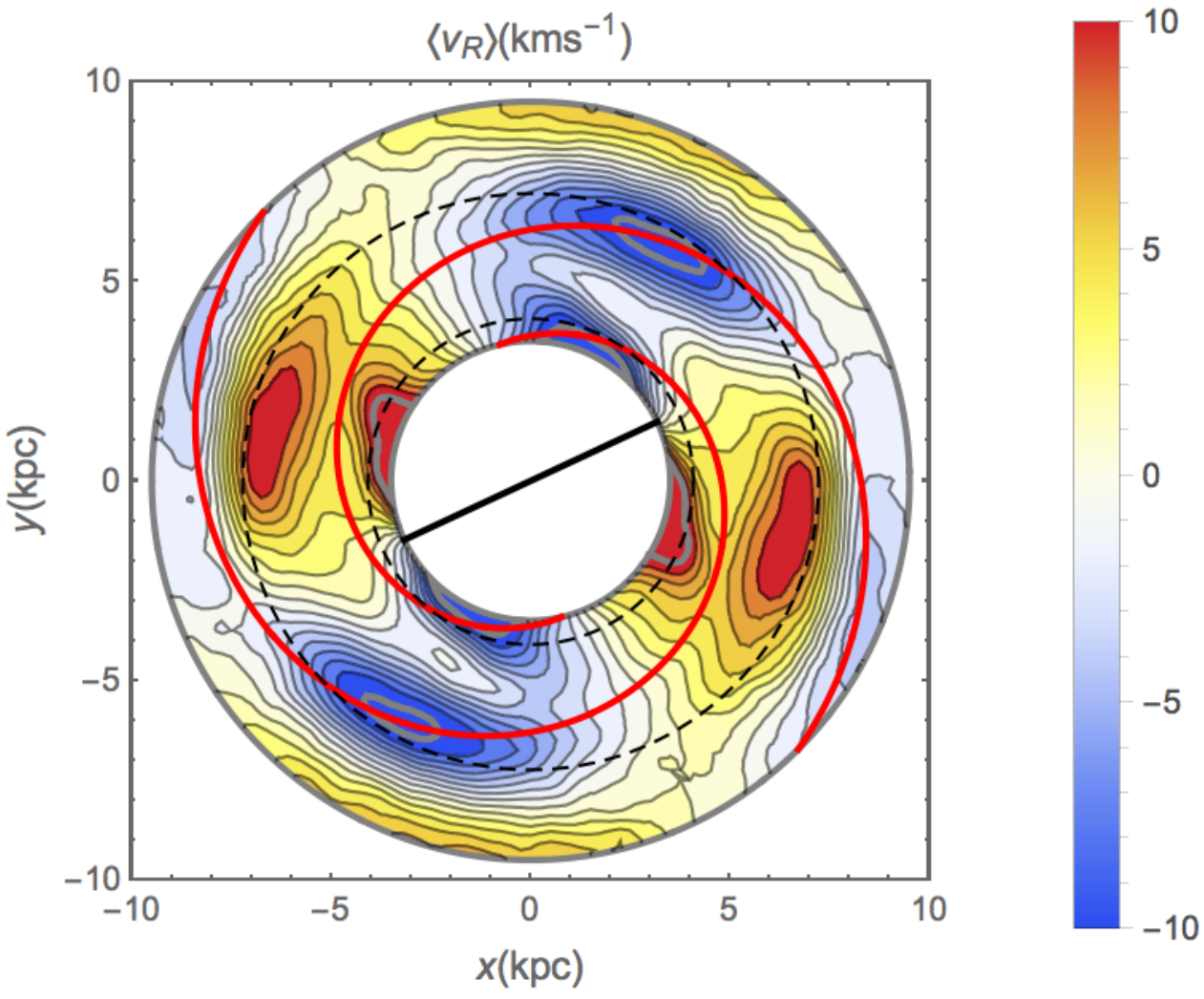}
  \includegraphics[width=0.32\textwidth]{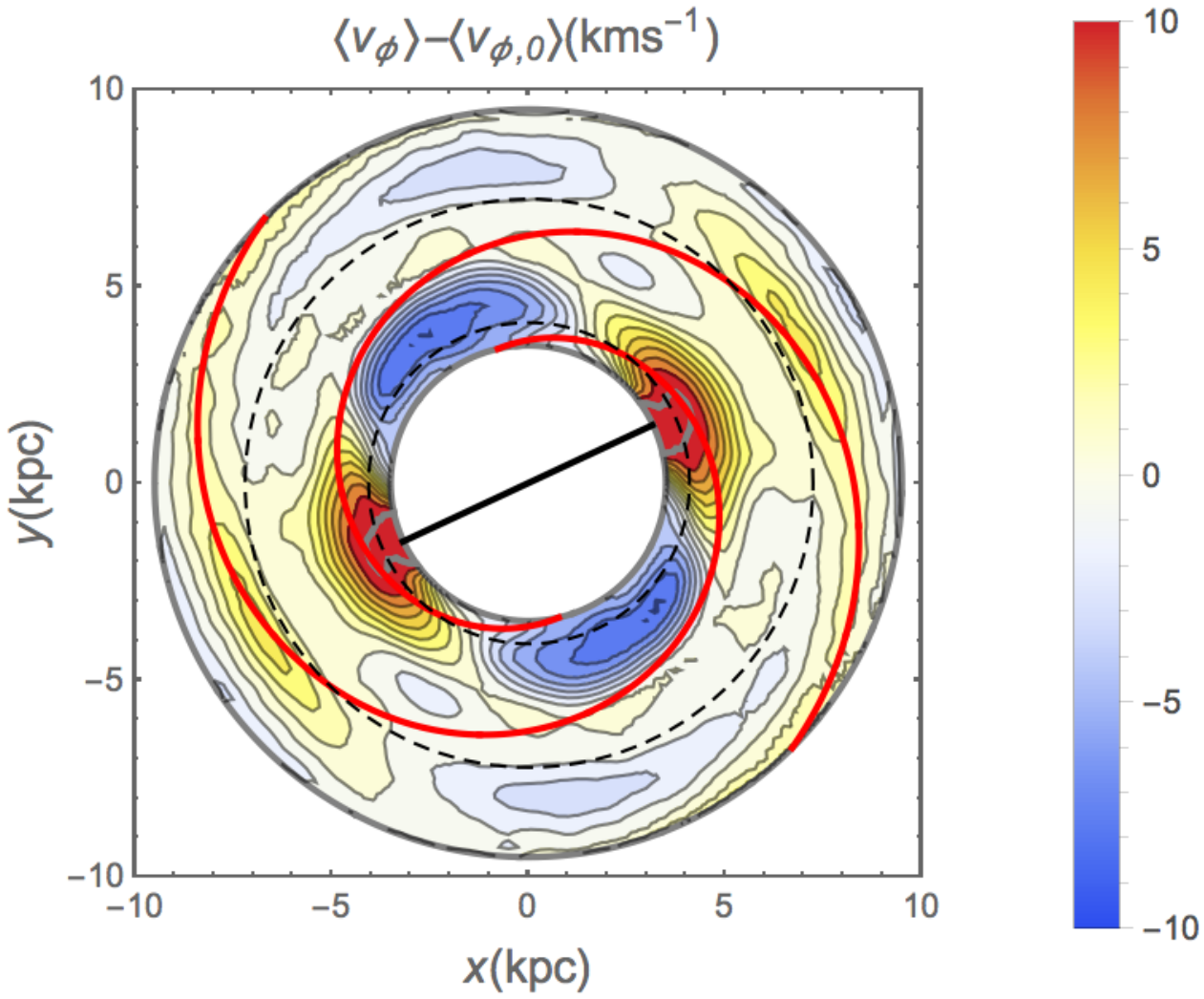}
  \includegraphics[width=0.32\textwidth]{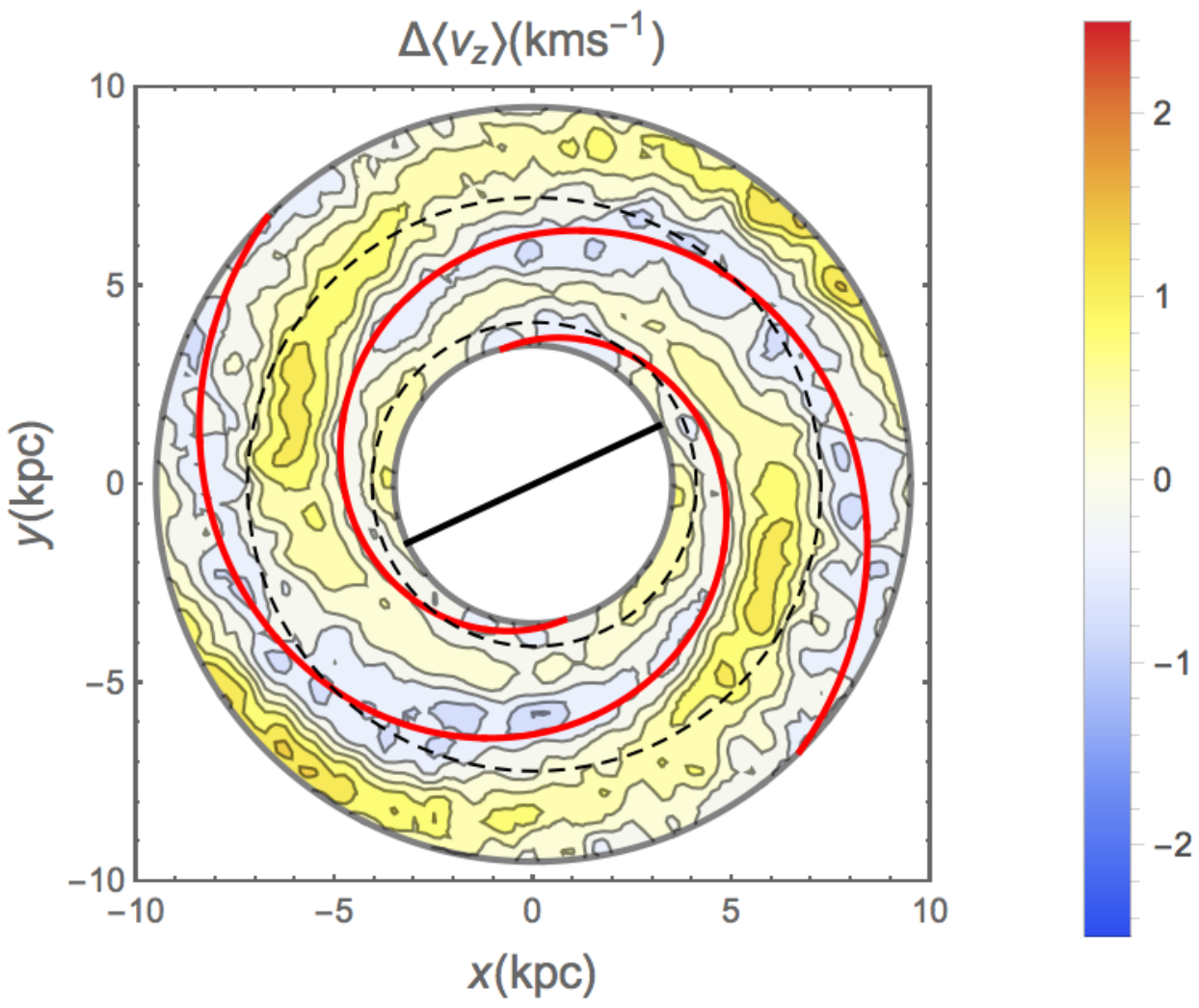}
  \caption{As in \Fig{fig:bar}, but for BS1.}
  \label{fig:barsp}
\end{figure*}
\begin{figure*}
  \centering
  \includegraphics[width=0.32\textwidth]{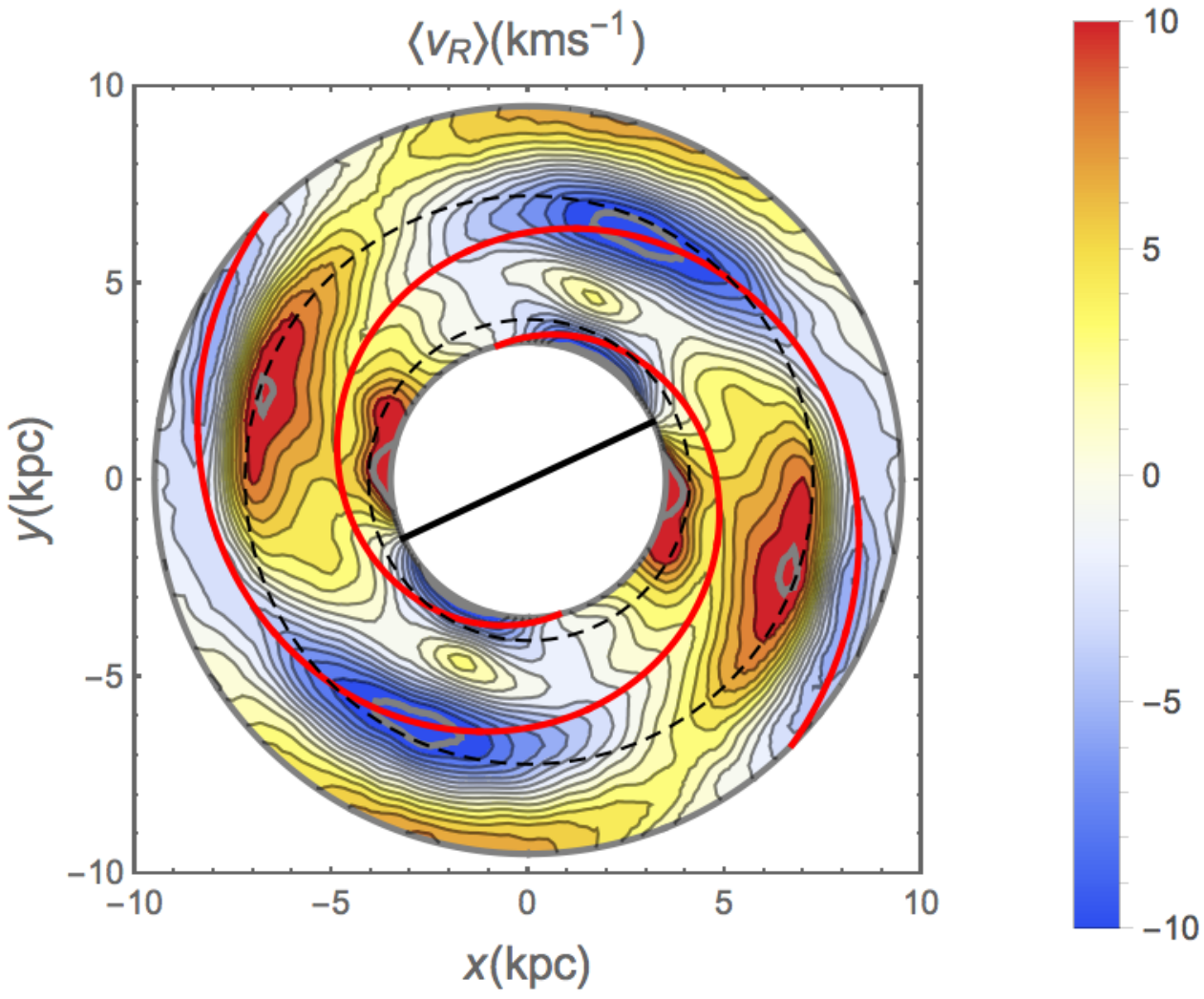}
  \includegraphics[width=0.32\textwidth]{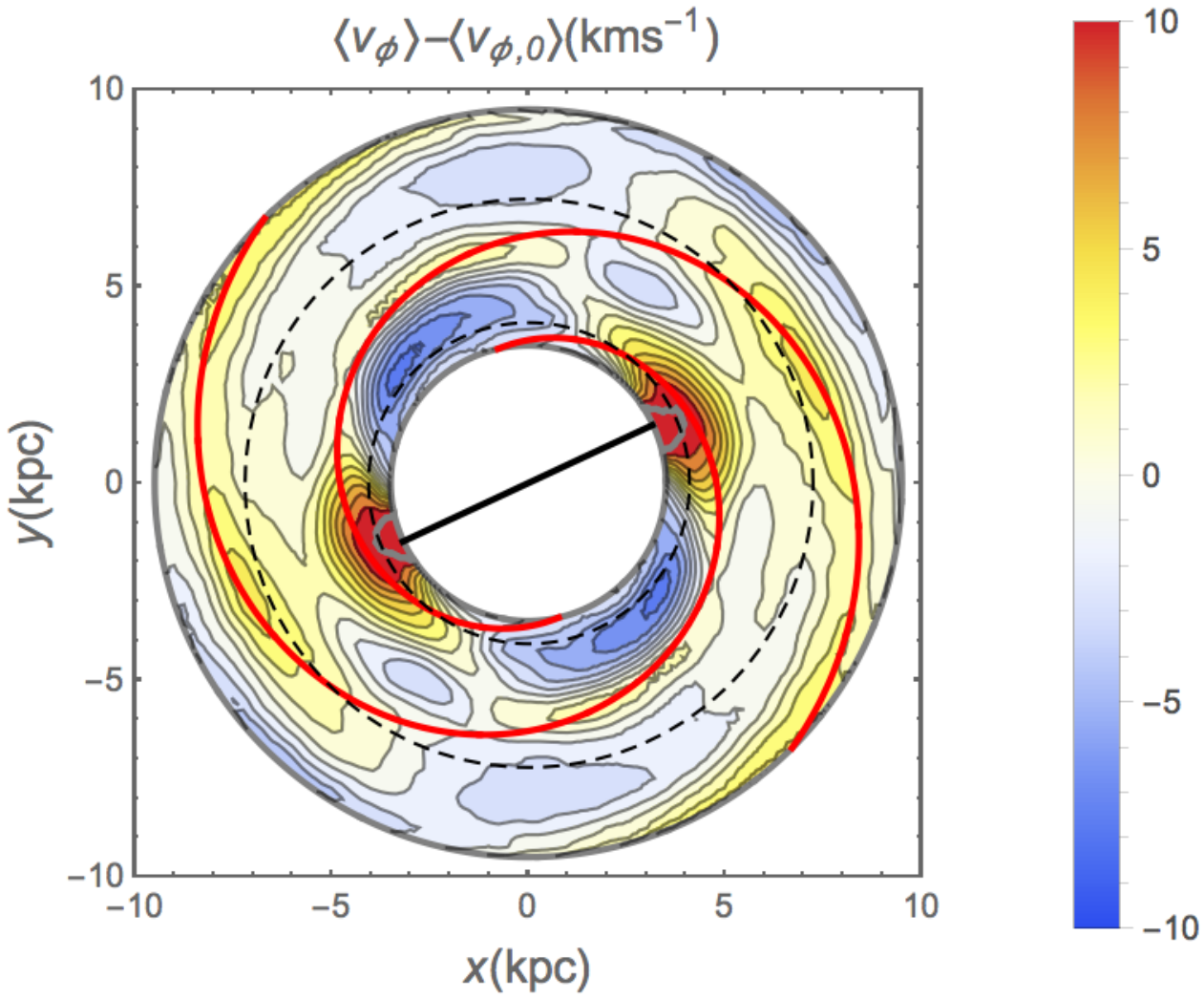}
  \includegraphics[width=0.32\textwidth]{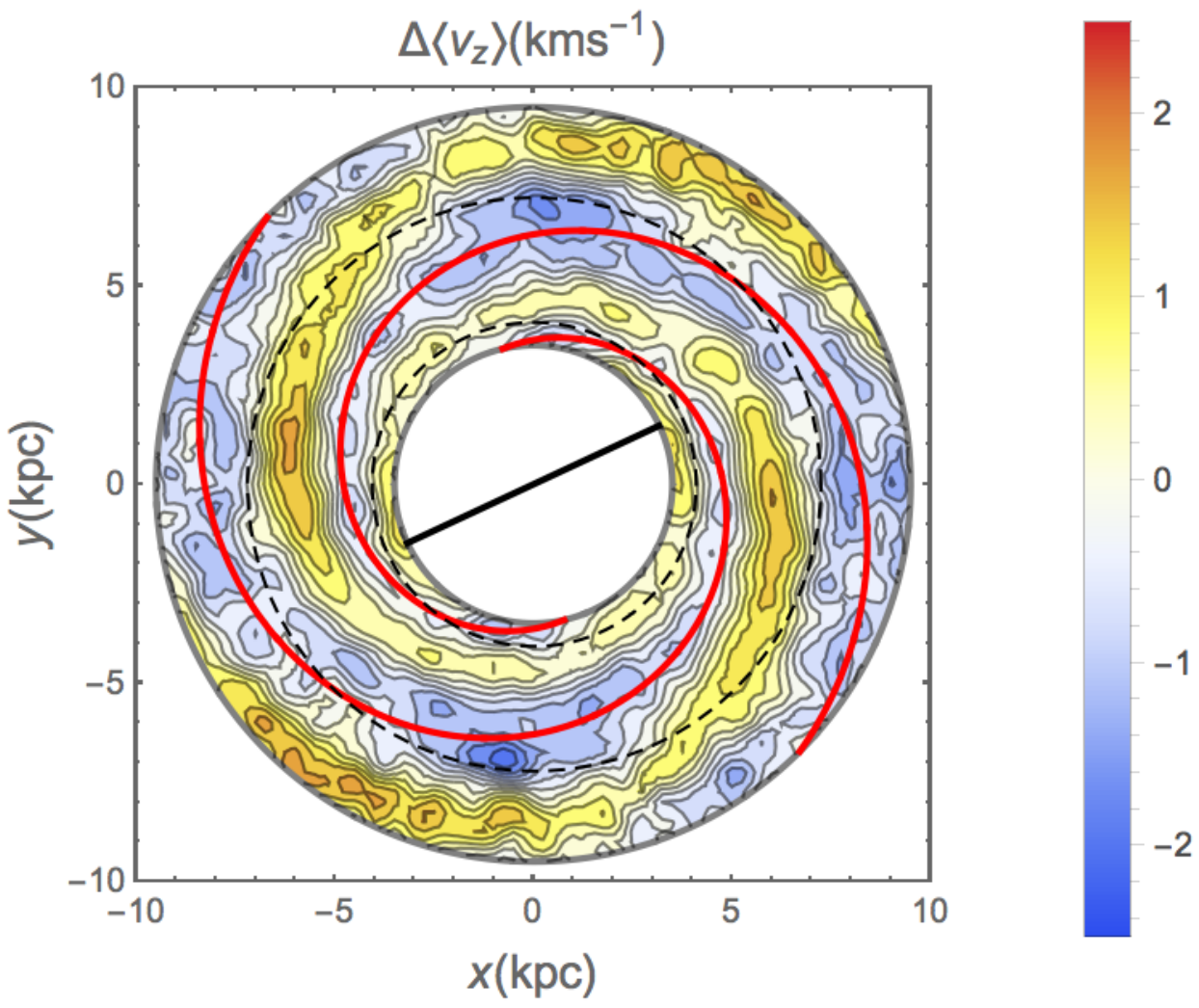}
  \caption{As in \Fig{fig:bar}, but for BS2.}
  \label{fig:barsp2}
\end{figure*}
We now focus on the main topic of this paper, namely the effects of
bar-spiral couplings on stellar kinematics. \Fig{fig:barsp} and
\Fig{fig:barsp2} represent the kinematical response of our simulations
in the case where both the bar and the spiral arms are present (BS1
and BS2 respectively). The first glance at these figures shows how the
bar seems to dominate the horizontal motions ($v_R$ and $v_\phi$), in
line with the analysis of the power spectrum, while the vertical
motions have the shape of spiral arms. However, concerning vertical
motions, comparing \Fig{fig:barsp} with \Fig{fig:sp}, and
\Fig{fig:barsp2} with \Fig{fig:sp2} we see that the effect of the bar
in $\Delta \langle v_z \rangle$ is to enhance the amplitude of the
breathing mode and to shift the position of the `vertical kinematic
spiral arms' w.r.t. the locus of the spiral arms. In particular,
while in \Figs{fig:sp}{fig:sp2} the locus of the spiral arms coincides
with the passage from $\dvz>0$ to $\dvz<0$, in
\Figs{fig:barsp}{fig:barsp2} the locus of the arms corresponds with
minima of $\dvz$ (i.e. regions where the particles move on average
towards the Galactic plane).

\begin{figure*}
  \centering
  \includegraphics[width=0.32\textwidth]{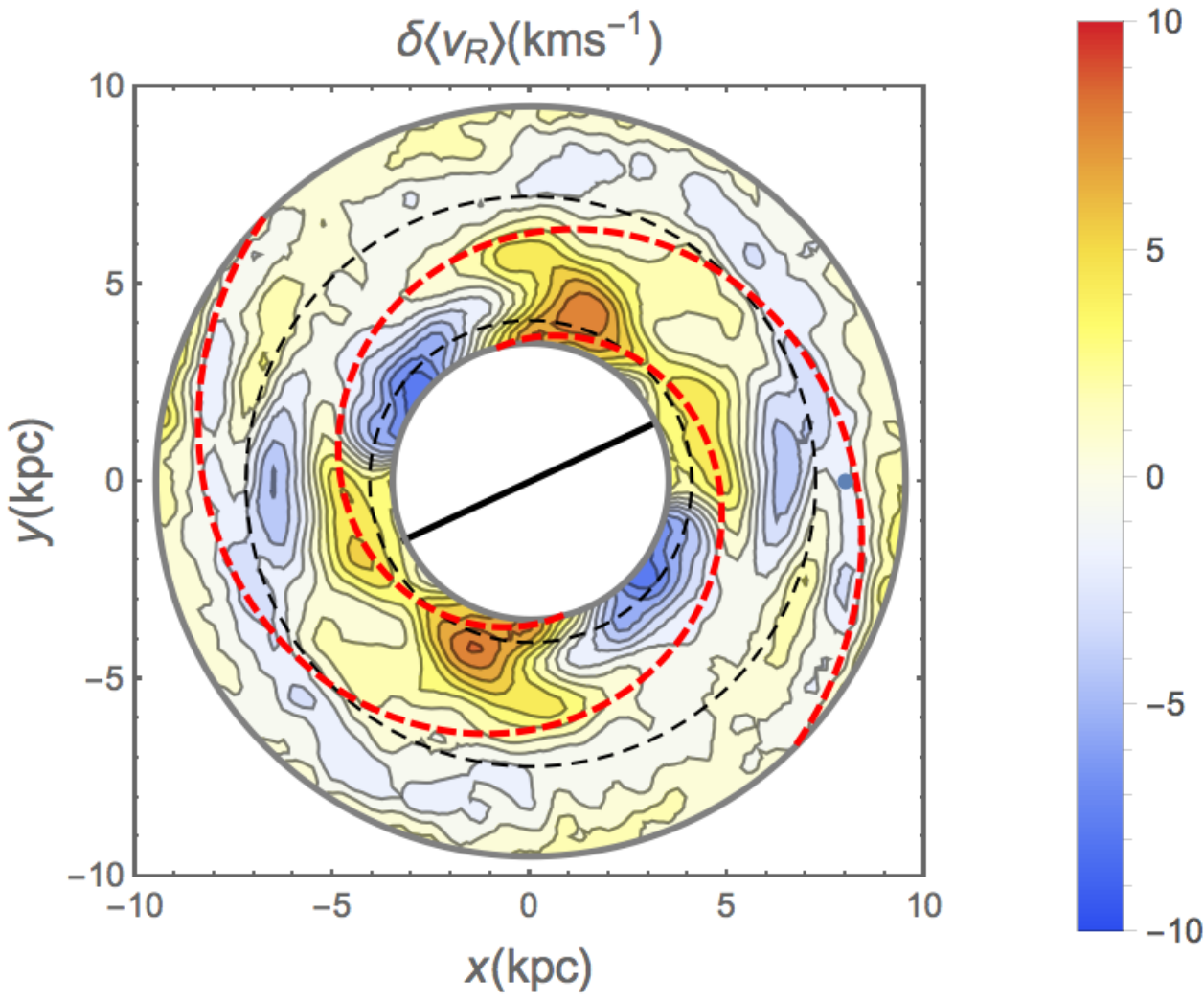}
  \includegraphics[width=0.32\textwidth]{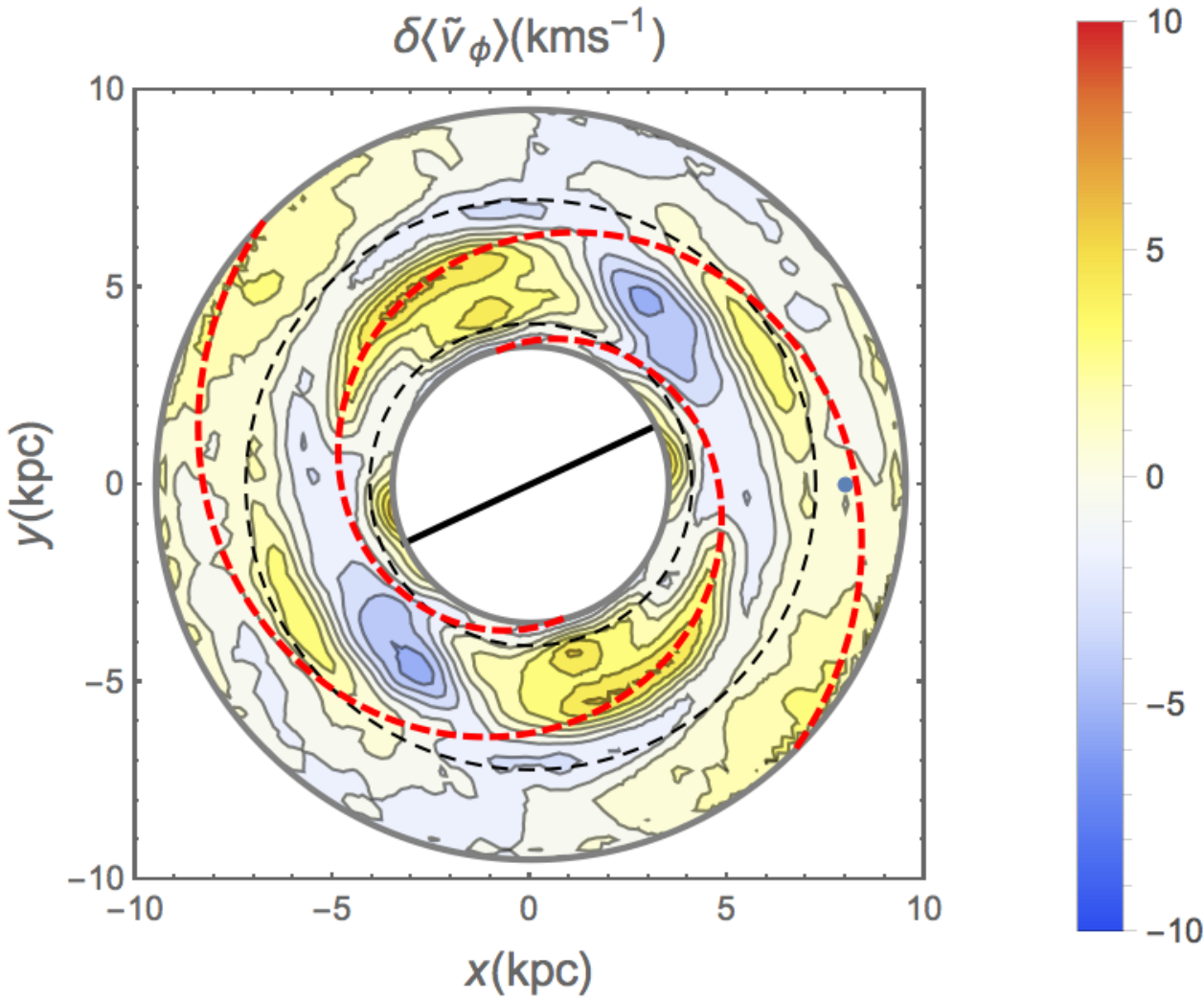}
  \includegraphics[width=0.32\textwidth]{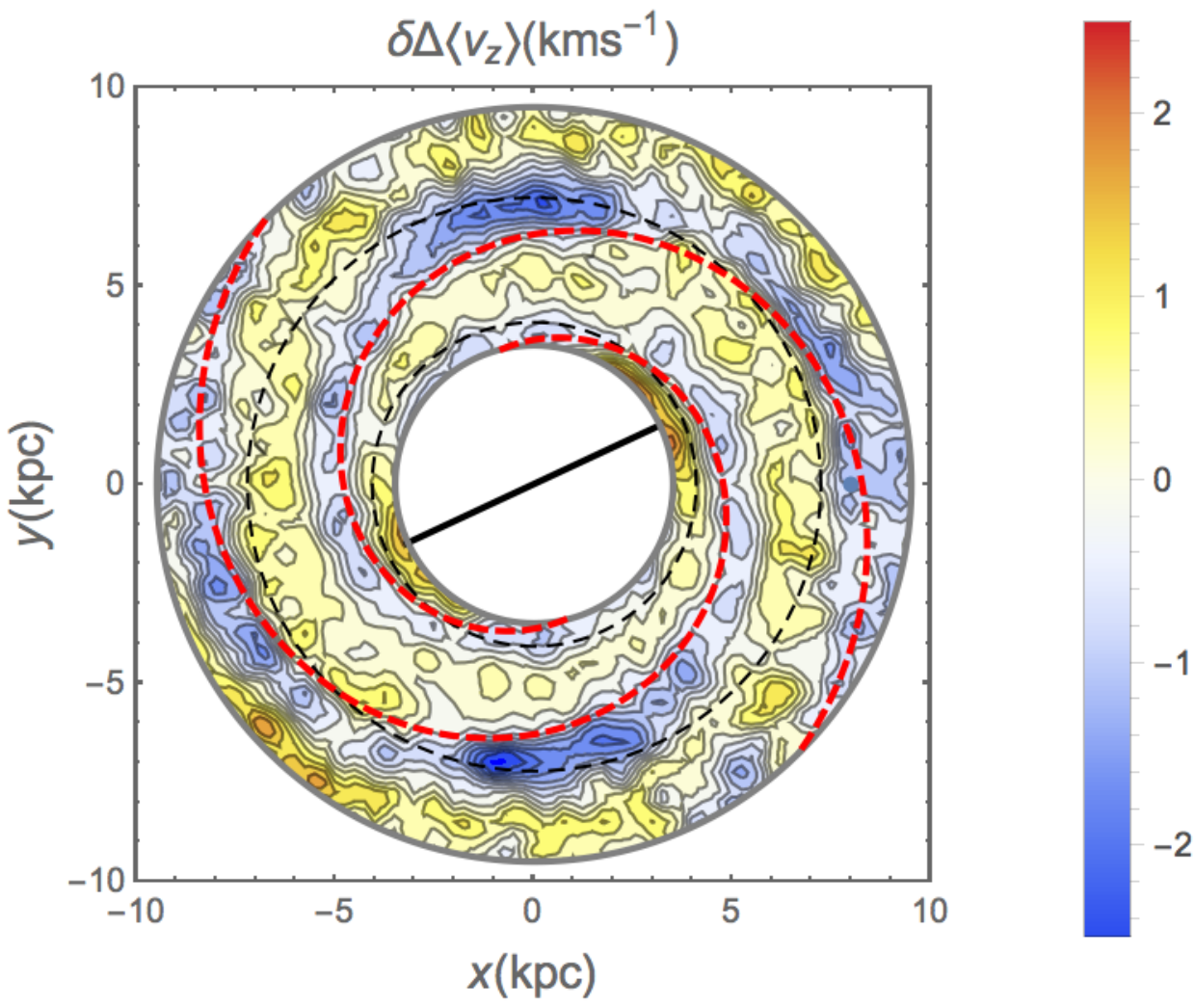}
  \caption{Comparing the kinematic effects of simulations B+S2 with
    simulation BS2 at $t=6\Gyr$. Left: $\delta\avvR$ in the
    $x$~vs.~$y$ plane. Center: $\delta\avvphi$ in the $x$~vs.~$y$
    plane. Right: $\delta\dvz$ in the $x$~vs.~$y$ plane. Bin sizes
    $0.25\Kpc$. See Eq.~\ref{q} for the definition of the plotted
    quantities.}
  \label{fig:deltabarsp2}
\end{figure*}
%% \begin{figure*}
%%   \centering
%%   \includegraphics[width=0.32\textwidth]{eta_vR_barsp25deg.pdf}
%%   \includegraphics[width=0.32\textwidth]{eta_vphi_barsp25deg.pdf}
%%   \includegraphics[width=0.32\textwidth]{eta_vz_barsp25deg.pdf}
%%   \caption{Simulations B, S1, and BS1 at $t=6\Gyr$. Left: $\eta$ for
%%     $\avvR$ in the $x$~vs.~$y$ plane. Center: $\eta$ for $\avvphi$ in
%%     the $x$~vs.~$y$ plane. Right: $\eta$ for $\dvz$ in the $x$~vs.~$y$
%%     plane. Bin sizes $0.25\Kpc$.}
%%   \label{fig:etabarsp}
%% \end{figure*}
%% \begin{figure*}
%%   \centering
%%   \includegraphics[width=0.32\textwidth]{eta_vR_barsp225deg.pdf}
%%   \includegraphics[width=0.32\textwidth]{eta_vphi_barsp225deg.pdf}
%%   \includegraphics[width=0.32\textwidth]{eta_vz_barsp225deg.pdf}
%%   \caption{As in \Fig{fig:etabarsp} but for the simulations B, S2,
%%     and BS2.}
%%   \label{fig:etabarsp2}
%% \end{figure*}
It is at this point important to quantify the non-linear effects of
the coupling of the bar and spirals on mean motions by comparing the
kinematics of our coupled simulations with the linear combination of
the single effect of these two perturbers. To study this, we use the
quantities
%% \begin{equation}
%%   \eta[q]\equiv \frac{\qBS-\pare{\qB+\qS}}{|\qBS|},
%% \end{equation}
\begin{equation}
\label{q}
   \delta q\equiv \qBS-\pare{\qB+\qS},
\end{equation}
where the superscripts B, S, and BS refer to estimating the quantities
in the bar, spirals (reference or strong), and bar and spirals
simulations respectively, whilst the quantity $q$ will be,
respectively, $\avvR$, $\tavvphi\equiv\avvphi-\avvphiz$, and $\dvz$.
The $\delta$ quantities represent the kinematic difference between the
models where the coupling between the bar and the spiral arms is
present and the linear combination of the effects of the bar and
spiral arms alone. \Fig{fig:deltabarsp2} shows, from left to right,
the quantities $\delta\avvR$, $\delta\tavvphi$, $\delta\dvz$ for the
BS2 simulations.
%% We show in red the regions where $\eta[q]>1$, $\qBS>0$,
%% $(\qB+\qS)>0$, in blue where $\eta[q]<-1$ $\qBS<0$, $(\qB+\qS)<0$, in
%% green where $\eta[q]>1$ $\qBS>0$, $(\qB+\qS)<0$, and in yellow where
%% $\eta[q]<-1$, $\qBS<0$, $(\qB+\qS)>0$. 
The $\delta \langle v_R \rangle$ panel reveals that significant
non-linear effects due to the coupling are restricted only to few
regions of the Galactic plane, % in the case of the $v_R$ velocity,
especially at the tip of the bar, and at the OLR, where resonance
overlaps with low order resonances of the spirals are taking place.
%% The $v_\phi$ and $v_z$ cases are
%% different. The plots of $\eta[v_\phi]$ present large regions where
%% $|\eta[v_\phi]|>1$, especially just inside the OLR.

The $\dvz$ case is different. In this case, the regions where
$\delta\dvz$ has a similar amplitude to $\dvz$ in the spiral arms case
extend in a large area of the Galactic plane, and have a form that
resembles spiral arms. In the arms regions $\delta\dvz$ is negative
(i.e. there is a surplus `compression' of the breathing modes),
while it is positive in the interarm regions (i.e. there is a surplus
`rarefaction' of the breathing mode). Notice that this implies, in
some regions, a change in the sign of $\dvz$, i.e. the passage from a
compression to a rarefaction breathing mode. This happens for example
just outside the loci of the spiral arms.

\begin{figure}
   \centering
   \includegraphics[width=\columnwidth]{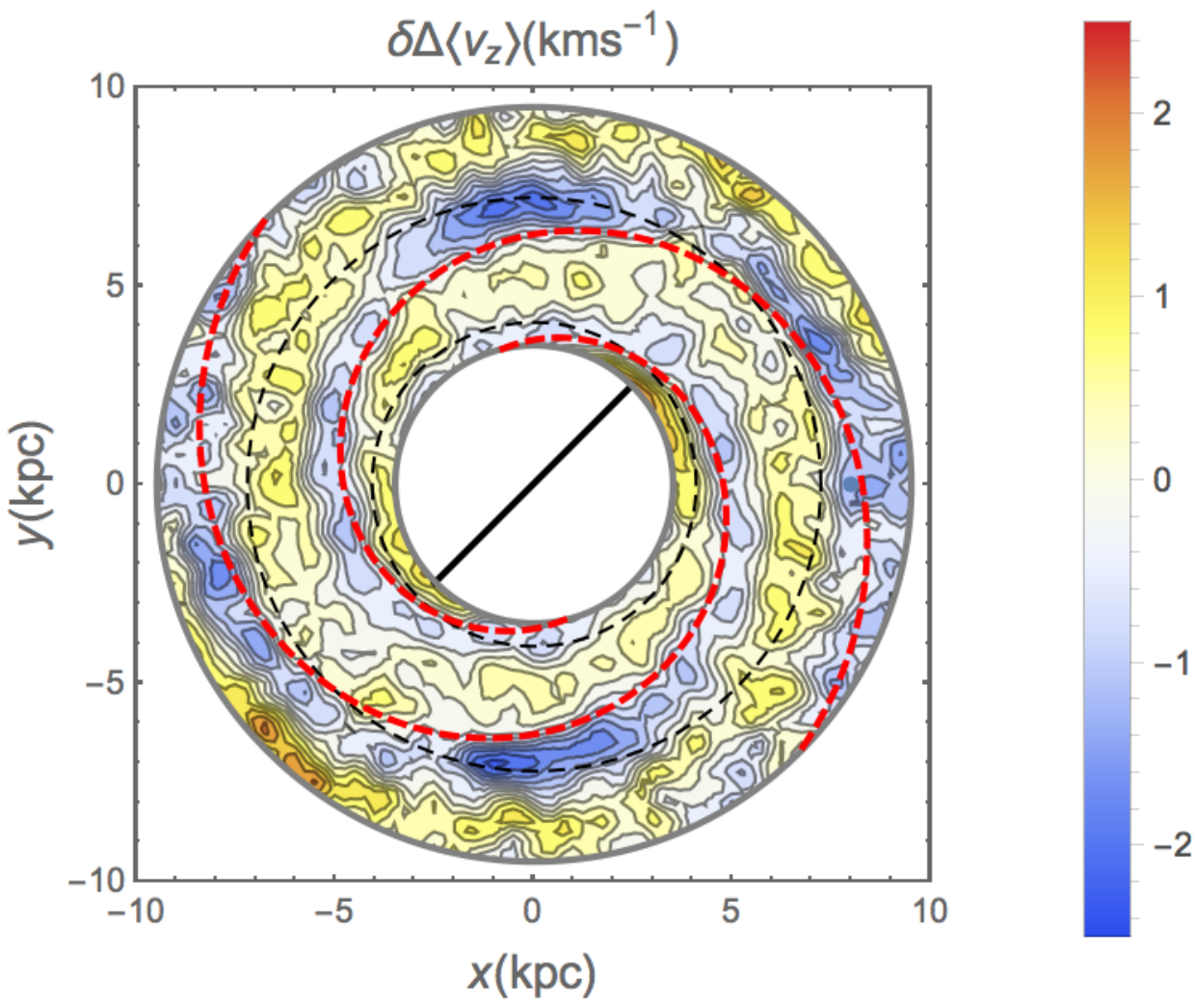}
   \caption{As in \Fig{fig:deltabarsp2}, right panel, but for the
     simulation B2S2 compared to B2+S2.}
   \label{fig:deltaDvz45}
\end{figure}

This particular configuration of $\delta\dvz$ is not specific to the
particular bar and spiral arms orientation: in \Fig{fig:deltaDvz45} we
have the same plot for the simulation B2S2, where the bar long axis is
oriented at $\phi=45\degr$ from the Sun, and we still have
$\delta\dvz<0$ on the arms and $\delta\dvz>0$ in the interarm regions.
This behaviour of $\delta\dvz$ is a major new result, which could help
explain from non-axisymmetries alone the amplitude of the observed
breathing mode in the extended Solar neighbourhood, which will be
quantified more precisely with forthcoming Gaia data. We however note
that the simulations presented here never reach the amplitude reported
by the current observations
\citep[][]{Widrow2012,Williams2013,Carlin2013} which, at least far
from the Galactic plane, can even rise to $|\avvz|\sim
15\kmsec$. However, the bulk of particles that we study in our
simulations is closer to the Galactic plane ($|z|<0.3\Kpc$), where the
observed gradient is of the order of $\sim 10^{-2}\kmsec\pc^{-1}$,
thereby reaching $|v_z|\sim 3\kmsec$ at $z=0.3\Kpc$. Such amplitudes
are almost twice as high as our BS2 values and could probably be
reproduced using other models of spiral arms with a stronger vertical
force (for example, when the scale height $\hs$ is smaller). In any
case, the non-linear enhancement of the breathing modes in the coupled
case is a non-negligible effect to take into account in future
modelling, and will have to be understood theoretically by coupling
two perturbers in analytical models of the type developed in M16.
%% \begin{figure*}
%%   \centering
%%   \includegraphics[width=0.32\textwidth]{eta_vR_barsp245deg.pdf}
%%   \includegraphics[width=0.32\textwidth]{eta_vphi_barsp245deg.pdf}
%%   \includegraphics[width=0.32\textwidth]{eta_vz_barsp245deg.pdf}
%%   \caption{As in \Fig{fig:etabarsp} but for the simulations B2, S2,
%%     and B2S2.}
%%   \label{fig:etabar2sp2}
%% \end{figure*}

\section{Radial migration}\label{sect:radm}
\begin{figure*}
  \centering
  \includegraphics[width=0.32\textwidth]{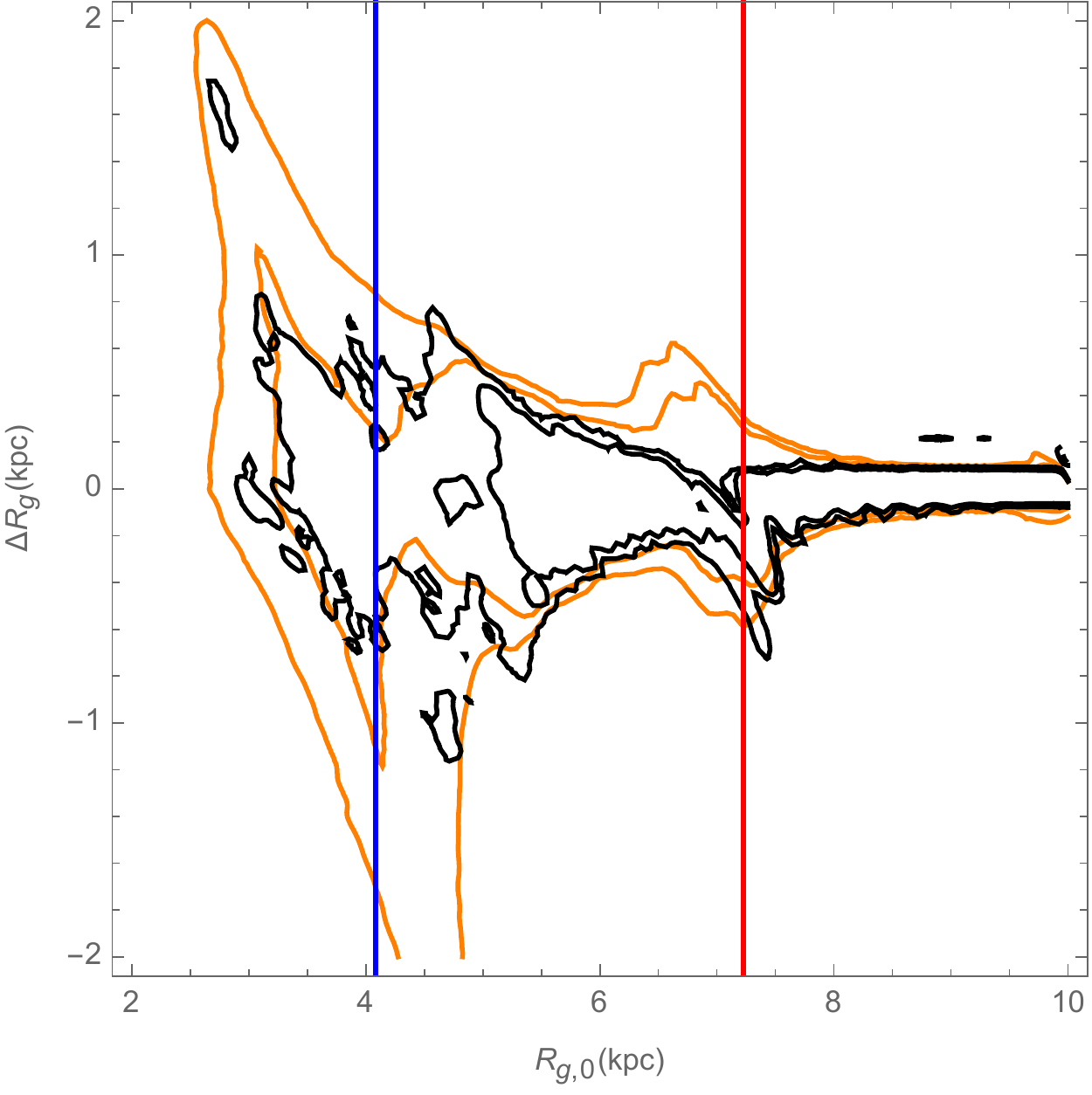}
  \includegraphics[width=0.32\textwidth]{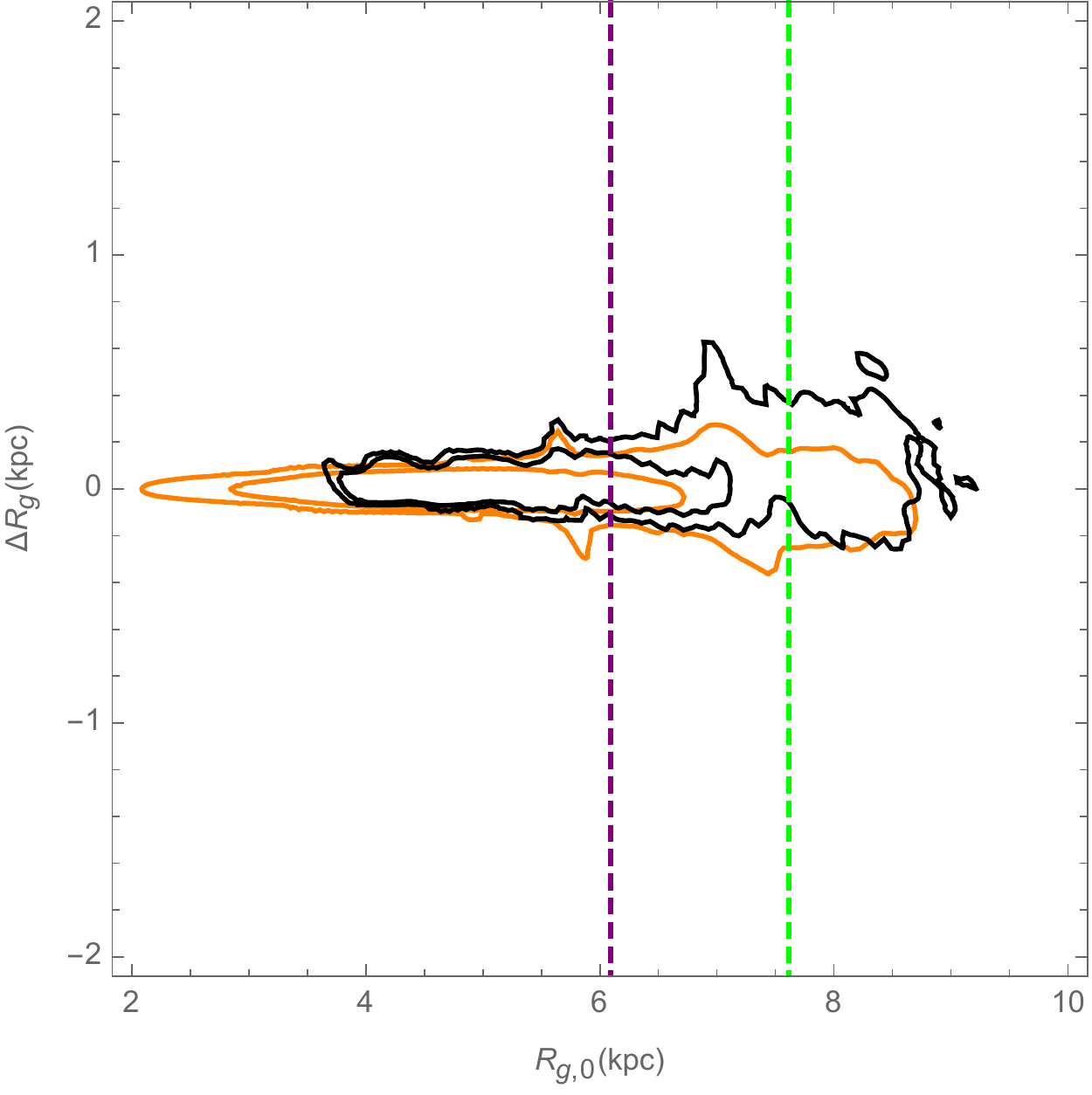}
  \includegraphics[width=0.32\textwidth]{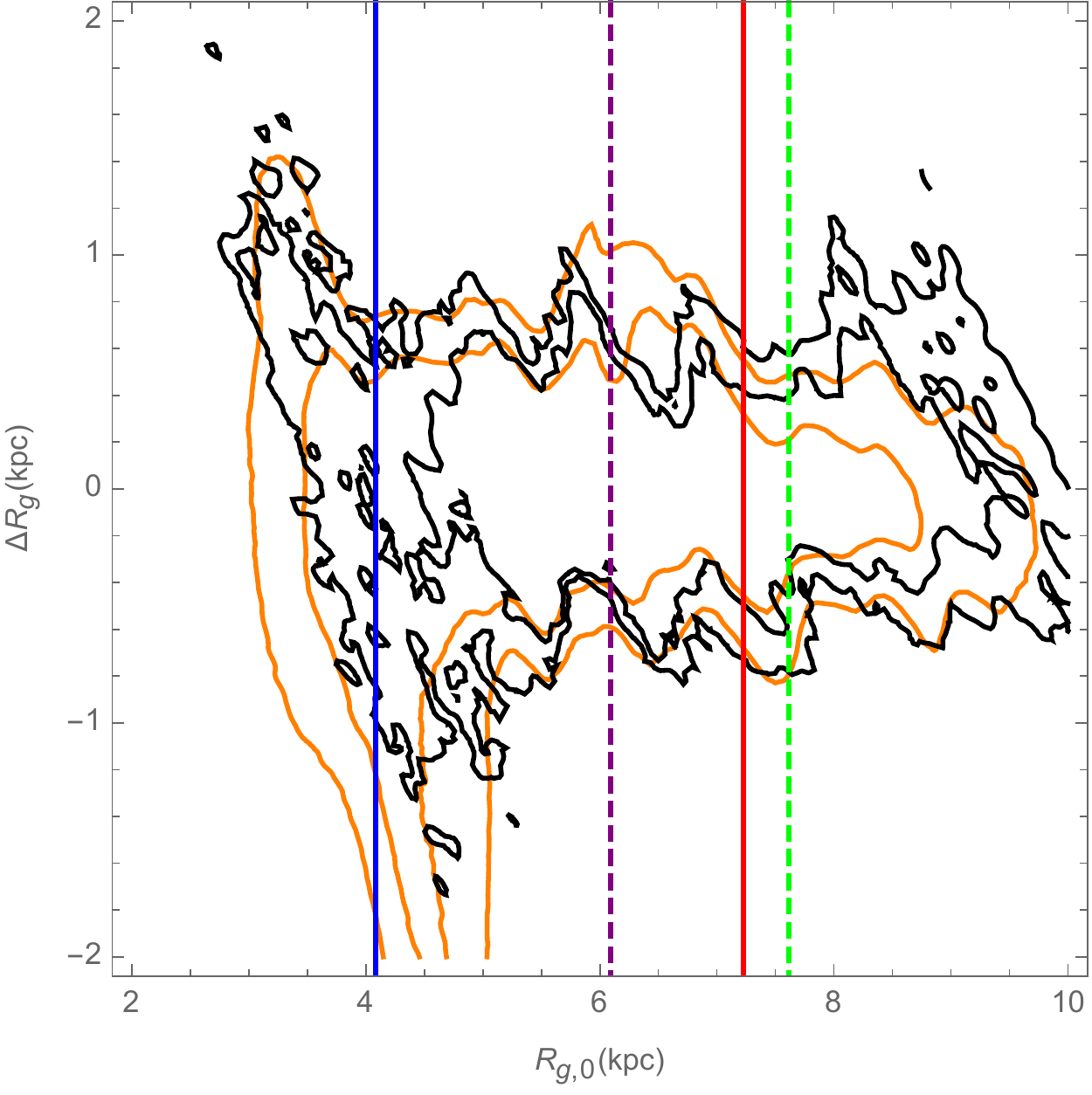}
  \caption{Distribution of the particles in our simulations in the
    space $\Rgz$ vs. $\Delta\Rg$, where $\Rgz$ is the guiding radius
    estimated $t=3\Gyr$, and $\Delta\Rg=\Rg-\Rgz$ with $\Rg$ estimated
    at $t=6\Gyr$. The black contours represent particles that have
    eccentricity $e<0.01$ at $t=6\Gyr$, while the orange contours all
    the particles. The contours include respectively 68 per cent and
    90 per cent of the particles. Left panel: B simulation. Central
    panel: S2 simulation. Right panel: BS2 simulation. The vertical
    lines represent the position of the resonances: the solid ones
    with the bar and the dashed ones with the spiral arms. In
    particular, the blue and red lines are the bar's corotation and
    OLR s respectively, and the violet and green lines the $\kappa :
    (\Omega-\Omegas) = 3:1$, and $\kappa : (\Omega-\Omegas) = 4:1$.}
  \label{fig:radm}
\end{figure*}
Of particular importance is the effect of non-axisymmetries of the
Galactic disc on the chemical enrichment of the stars and the chemical
evolution of the Galactic disc. The distribution of chemical
abundances and ages in the Solar neighbourhood seems to be
incompatible with a simple model where the stars are born from
progressively metal enriched cold gas at a given radius, and which
does not change except for the oscillations due to the eccentricity of
their orbits \citep[e.g.][]{SellwoodBinney2002}. The exact amount of
so-called `radial migration" needed in chemical evolution models to
explain observations is nevertheless subject to much debate
\citep{MinchevMartig2013,KubrykPrantzos2015,Haywood2016}.

This mechanism is linked to the non-axisymmetric structures of the
Milky Way moving the guiding centre of the stars' orbits around the
disc via resonant trapping, without changing their
eccentricities. This is also referred to as `churning', to not be
confused with `blurring' corresponding to the increase in velocity
dispersions which can also make stars span wider radii, but without
changing their guiding radius. The radial migration mechanism
originally proposed by \cite{SellwoodBinney2002} involves transient
spirals, which trap stars on horseshoe orbits close to their
corotation before fading away, and thus prevents stars from returning
to their initial guiding radius. This effect has been shown to be
significantly increased in the presence of multiple spiral patterns of
different angular speed, or in the case of the coupling of the bar and
spiral arms \citep{MinchevFamaey2010}. This increase has been
attributed to resonance overlaps \citep{Chirikov1979}, i.e. that an
orbit is trapped first by one resonance with one pattern, and then by
another one with the other pattern, with the times of transition
between resonantly trapped families varying erratically. This chaotic
behaviour can enhance churning indeed, but can at the same time cause
some blurring of the orbits.

In this Section we reanalyse the pure effect of resonance overlap on
the amount of radial migration in our simulations for which amplitudes
are {\it not} varying once the perturbations have settled, in the
spirit of \cite{MinchevFamaey2010}, but this time in 3D and with
low-order resonance overlaps. In \Fig{fig:radm} we show the change of
guiding center radii induced by the bar and spiral arms, for a
selection of particles with low eccentricity ($e<0.01$) at the end of
our simulations. In this way we aim to separate the particles that
actually migrated (`churning') from the `blurring' of the disc
\citep[see,][]{SellwoodBinney2002}, which might affect all particles
in the disc (orange contours in \Fig{fig:radm}). The initial guiding
center radius $\Rgz$ is computed at the moment when spiral arms and
the bar are completely grown ($t=3\Gyr$), while the final one is at
the end of the simulation ($t=6\Gyr$). The guiding centers $\Rg$ are
found solving the equation $\Rg^2\Omega(\Rg)=L_z$ for each particle
with angular momentum $L_z$. The (epicyclic) eccentricity $e$ is
estimated \citep[e.g.][]{Dehnen1999} using
\begin{equation}\label{eq:ecc}
e=\sqrt{\frac{v_R^2+\kappa^2(\Rg)(R-\Rg)^2}{\kappa^2(\Rg)\Rg^2}}.
\end{equation}
The percentage of particles on almost circular orbits (i.e. with
$e<0.01$) is $\sim 0.2$ per cent in our simulations.

\Fig{fig:radm} shows that, while in the case of the single
perturbations radial migration is very limited as expected for
non-varying amplitudes, the coupled effects of bar and spiral arms, is
not simply the sum of their single effects. Rather, because of the
coupling, the stars migrate in the whole range of radii that we
consider, and the amplitude of the migration is strongly enhanced,
with 32 per cent of low-eccentricity orbits being transported without
heating on scales of at least $\sim 0.8\Kpc$, and 10 per cent more
than $1\Kpc$ in the outer disc.

This result is in qualitative agreement with the 2D study of
\cite{MinchevFamaey2010}, who showed that churning can be driven even
by static density waves provided resonance overlaps of multiple
patterns are present. With the parameters adopted here, it appears
clearly that this mechanism is indeed real, albeit more limited in
amplitude than what can be expected from scattering at corotation of
transient spirals \citep{SellwoodBinney2002}, and much more limited
than in the models of \cite{MinchevFamaey2010}. The differences in the
amplitude of the migration between most of these previous simulations
and the present ones should be ascribed to the different combination
of parameters and spiral arm models. Indeed, in
\cite{MinchevFamaey2010} the most efficient radial migration was
reached when the corotation of one pattern was overlapping with the
first-order Lindblad resonance of the other one (for instance if the
corotation of the spiral coincides with the OLR of the bar, or if the
spiral is four-armed and its 4:1 inner Lindblad resonance coincide
with the bar's OLR). In our case, the strongest overlap is between the
OLR of the bar and the 3:1 and 4:1 inner resonances of the two-armed
spiral, hence lower-order resonances than the corotation or inner
Lindblad resonance. The effect of these two low-order resonances of
the spiral on churning, as well as the effect of the bar's OLR, is
nevertheless clearly visible as distinct peaks in the rightmost panel
of Fig.~\ref{fig:radm}. Together with this, the spiral arms used in
the models by \cite{MinchevFamaey2010} (the `TWA spiral arms') are
stronger than the spiral arms used in our simulations in the regions
with $R<R_0$. For example, taking a maximum radial force of the
spirals as $1$ per cent of the force due to the axisymmetric
background at $R=R_0$ for both models, the force due to the TWA spiral
arms is $\sim 1.5$ ($\sim 2$) times larger than that of our spiral
arms at $R=6\Kpc$ ($R=5\Kpc$). Moreover, \cite{MinchevFamaey2010}
consider a range of amplitudes going from $0.5$ per cent to $3$ per
cent of the background force. Finally, the fact that our simulations
are 3D reduces the impact of the perturbing force for stars with large
oscillations outside of the Galactic plane, where the bar and spiral
arms radial and tangential forces are weaker (however, this is not the
case of the particles with $e<0.01$, which spend most of their orbits
close the Galactic plane). We thus note that, if the spiral pattern of
the Milky Way is a quasi-static density wave with realistic parameters
as chosen here, disc metallicity gradients would actually be
unaffected by churning; it would not be possible to explain the
age-metallicity relation at the solar vicinity, since differences in
metallicity of (at least) 0.5~dex, as those observed at all ages among
thin disc stars, would require migration to occur on several-kpc scale
($\sim 5-6\Kpc$) for a metallicity gradient of 0.1~dex/kpc, that is at
least a factor 5 higher than the scale of migration found in our
models. On the other hand, some simulations
\citep[e.g.][]{SellwoodCarlberg2014} tend to show that quasi-static
modes do not survive more than 10 rotation periods at corotation
($\sim 1$~Gyr), thereby also causing migration by scattering at
corotation \citep{SellwoodBinney2002}. Other models with co-rotating
spirals display even more drastic migration through this corotation
mechanism \citep{Kawata2014}. The exact nature of spirals is thus of
fundamental importance for theoretically quantifying radial migration,
as our models clearly show that, even though migration is indeed
happening also when spirals are kept static, the scale of migration is
actually extremely limited when adopting realistic parameters for the
Milky Way spiral arms.

In our simulations we also recognize a fraction of orbits that cross
the bar's OLR: in the simulation with only the bar we count $\sim 2$
per cent of them, with or without the cut in eccentricity. If we
restrict to the least eccentric orbits (as defined above) which have
$|\Rgz-\ROLR|<1\Kpc$ at $t=3\Gyr$ (the time when both bar and spiral
arms are fully grown and kept constant in amplitude afterwards), this
fraction increases to $\sim 10$ per cent.
%This is in agreement with
%\cite{LyndenBellKalnajs1972}, who predicted the absorption of angular
%momentum of stars on nearly circular orbits at the OLR (see
%\Fig{fig:radm}, left panel). 
Note that there is both absorption and emission in the coupled
case. However, at the quantitative level, our model is not really in
contradiction with the findings of \cite{Halle2015} who suggested that
the OLR limits the exchange of angular momentum between the inner and
outer disc. In fact, the mean amplitude of the excursions in guiding
radius of the particles that cross the OLR in our bar simulation is
moderate ($\sim 0.3\Kpc$ for all the particles, and $\sim 0.5\Kpc$ for
the least eccentric orbits), if compared to the rms epicyclic
amplitude (for all stars) at $R=\ROLR$ ($\sim 1.2\Kpc$). Taking into
account that the OLR region has non-null thickness
\citep[e.g.][]{CeverinoKlypin2007}, makes the interpretation of these
results even more complicated.

\section{Predictions for deep spectroscopic surveys}\label{sect:surveys}
\begin{figure*}
  \centering
  \includegraphics[width=\columnwidth]{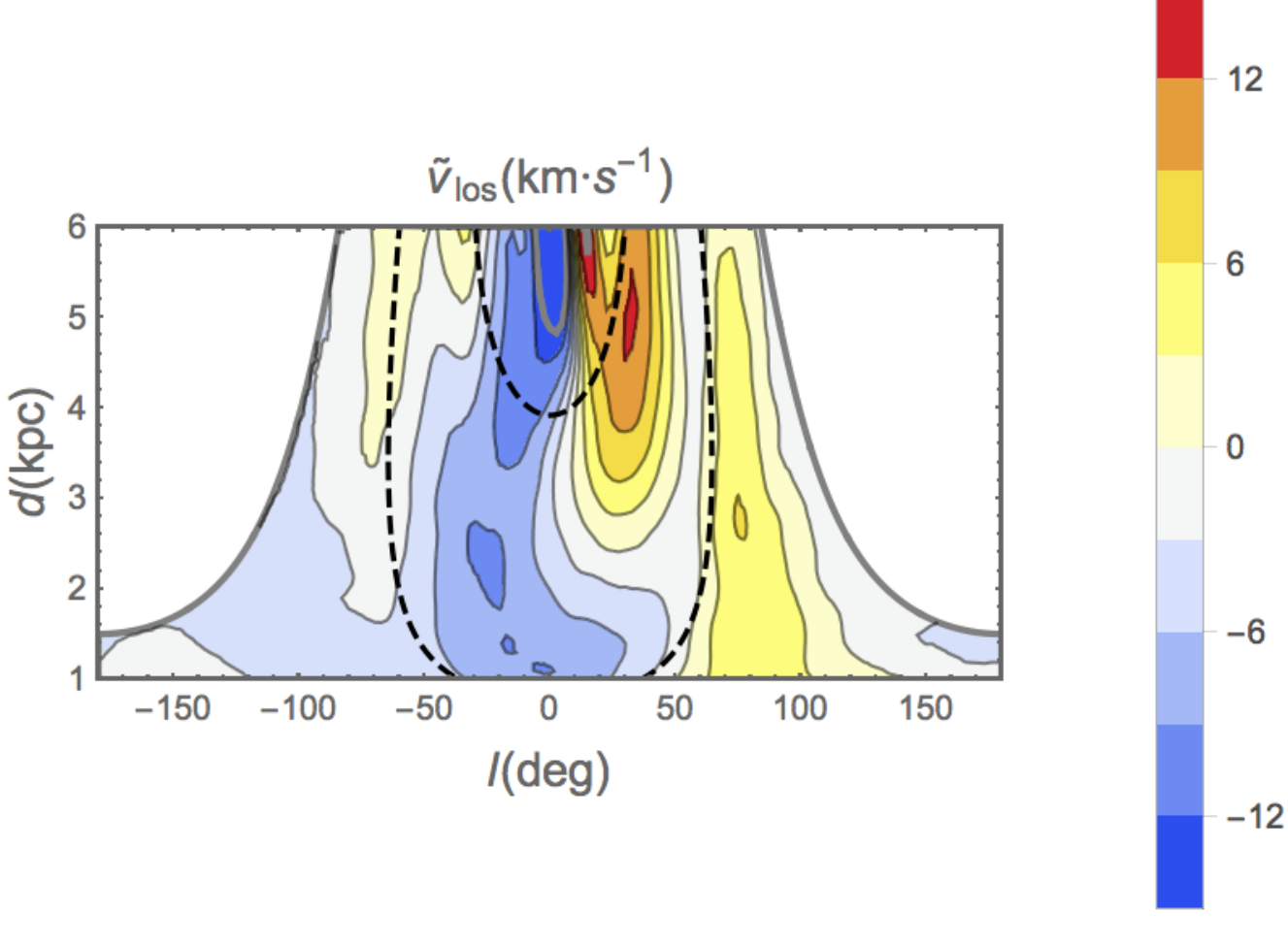}
  \includegraphics[width=\columnwidth]{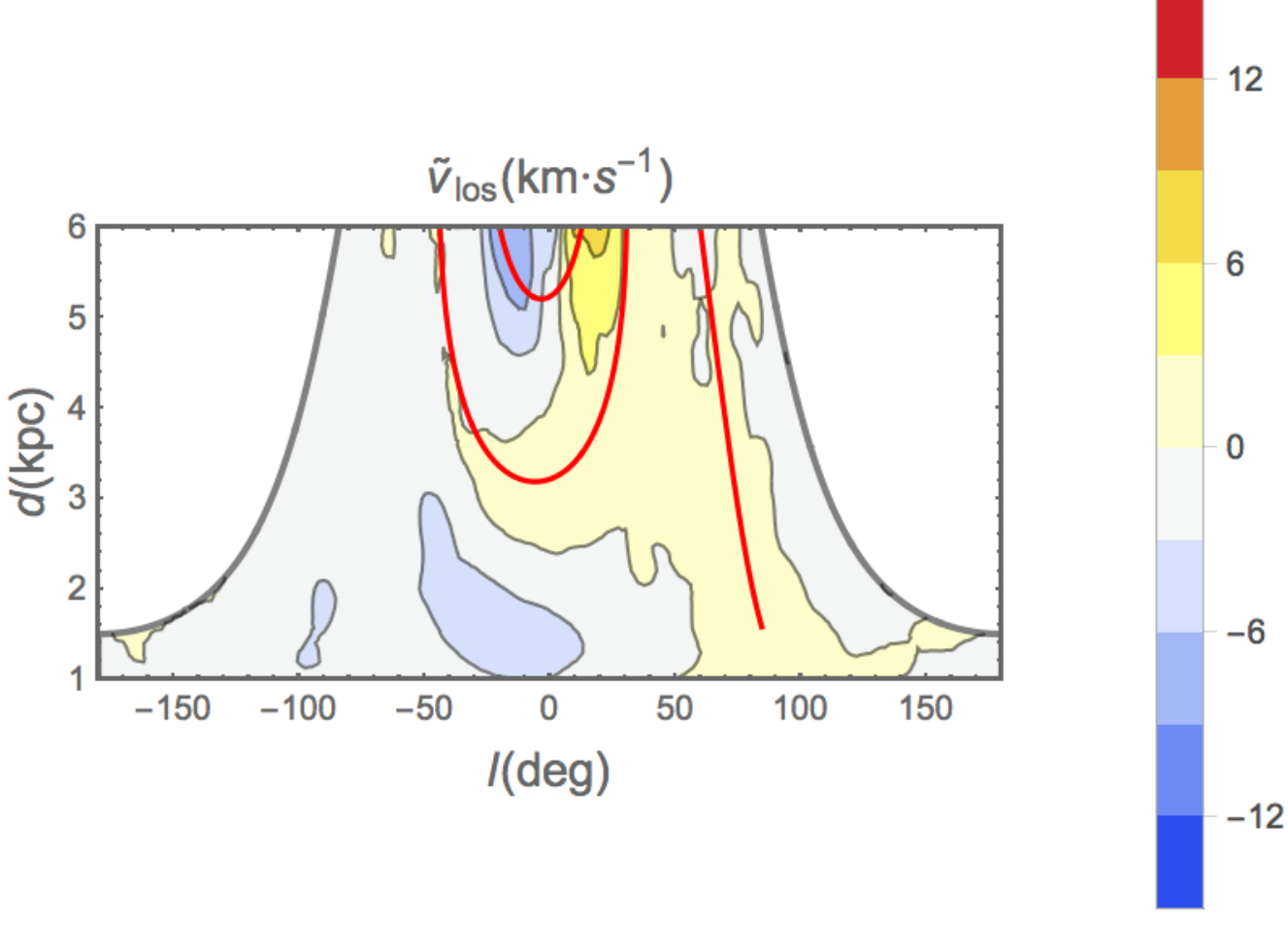}
  \includegraphics[width=\columnwidth]{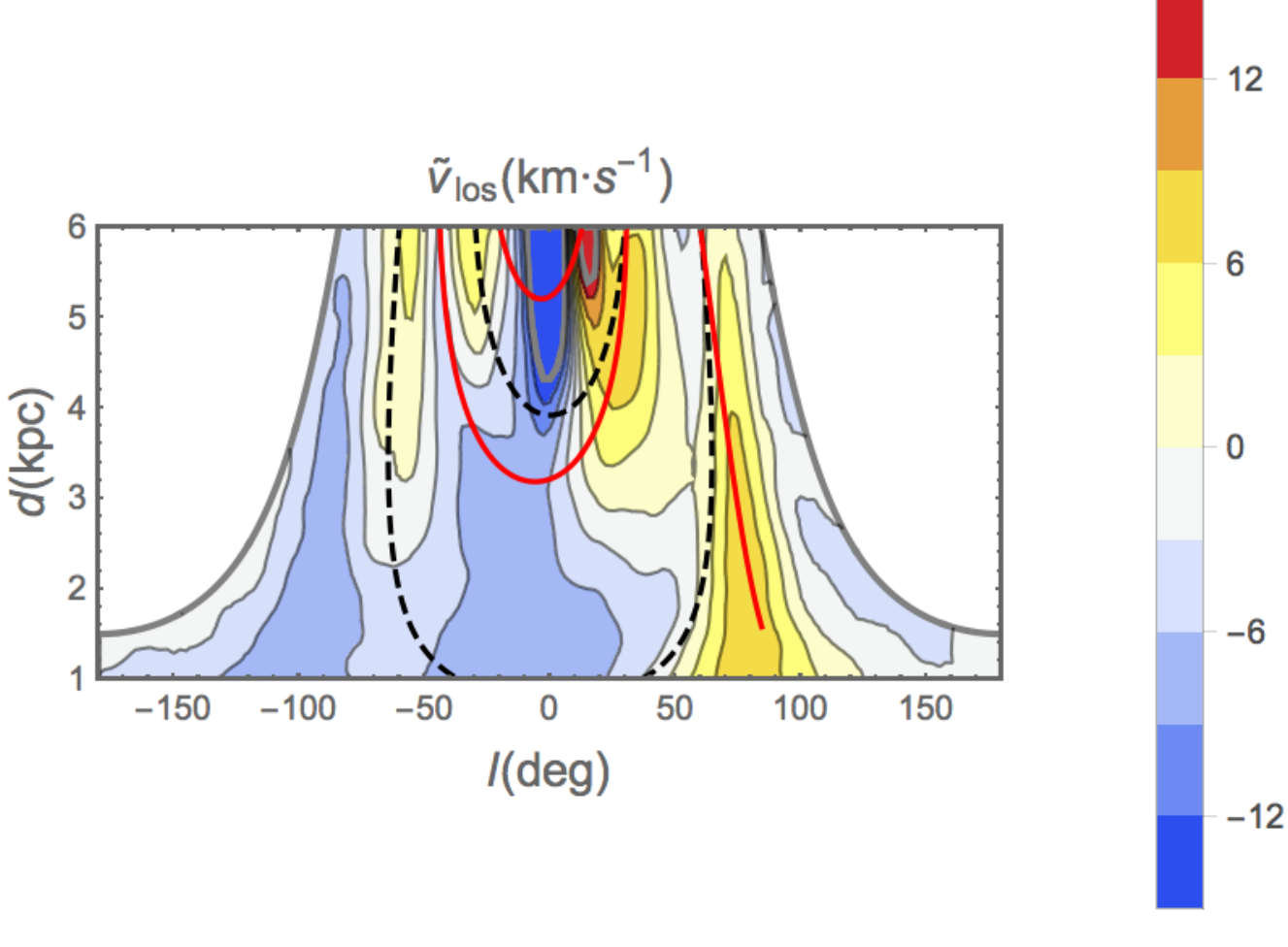}
  \includegraphics[width=\columnwidth]{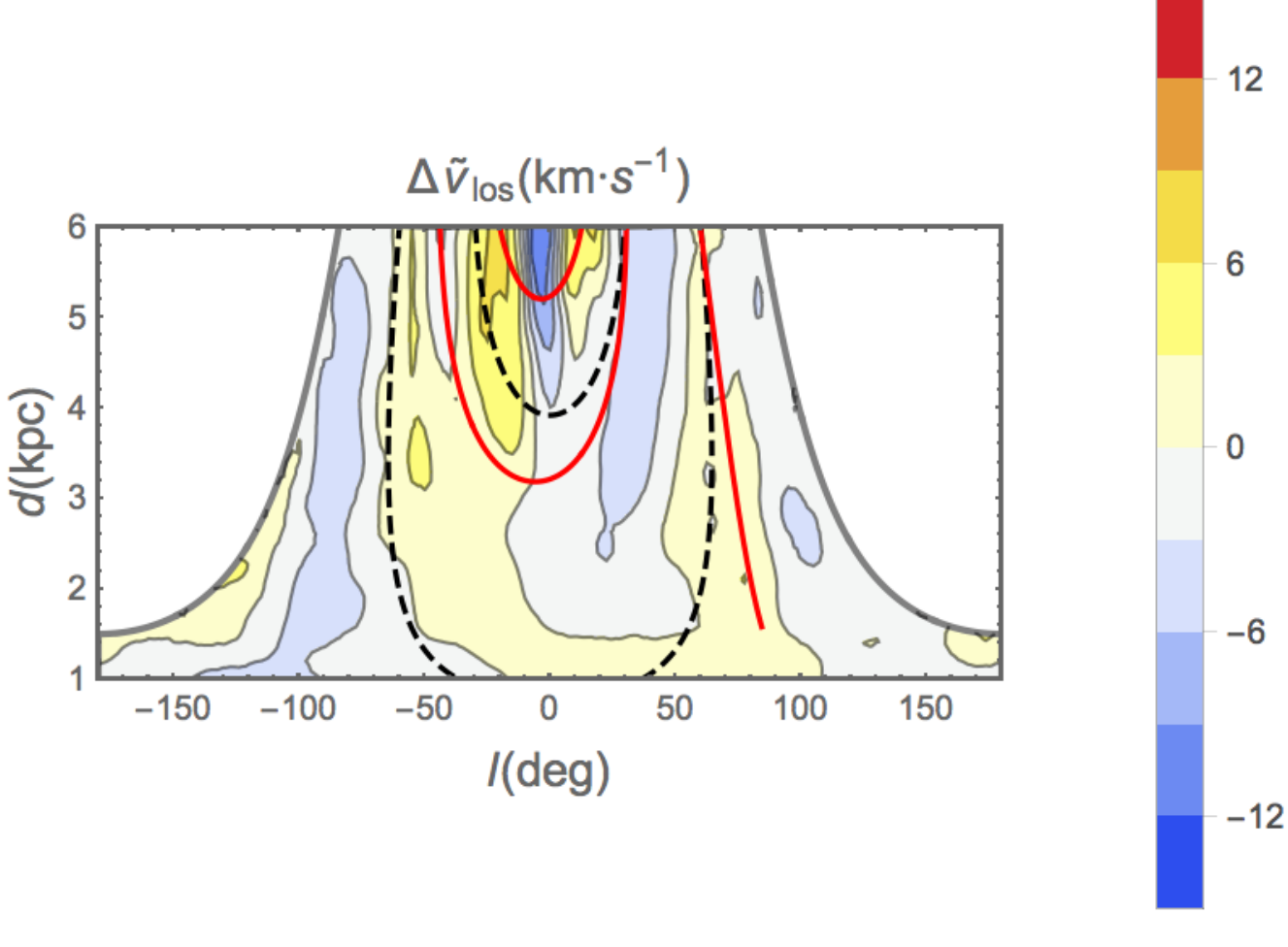}
  \caption{Peculiar velocity as a function of the Galactic longitude
    $l$, and distance $d$ from the Sun, computed for some simulations
    presented in this work. Top left panel: $\vlost$ for the B
    simulation. Top right panel: $\vlost$ for the S2
    simulation. Bottom left panel: $\vlost$ for the BS2
    simulation. Bottom right panel: $\Delta\vlost$, obtained
    subtracting $\vlost$ for BS2 and $\vlost$ for B. The particles
    used to compute these plots have $|b-2\degr|<1\degr$. The black
    dashed curves represent the corotation and the outer Lindblad
    resonance of the bar, the red curves the loci of the spiral arms.}
  \label{fig:vlos}
\end{figure*}
Ongoing and forthcoming spectroscopic surveys of the Galaxy will be
extremely useful to disentangle the effects of the bar and spirals in
the Galaxy. As we have seen here, the in-plane peculiar velocity power
spectrum is dominated by the bar, and hence makes it difficult to
constrain the effect of spirals. In this respect, constraining the
local vertical breathing mode might yield a lot of information on the
spirals, but this could be difficult to disentangle from other
external effects in the outer disc, such as bombarding of the disc by
small dark matter sub-halos \citep{Grand2016,Gomez2016}. In the inner
disc, obtaining exquisite constraints on the breathing mode in 3D
could on the other hand be difficult due to the absence of good enough
astrometric data, even with the advent of Gaia \citep[for this, we
  will need to wait for Theia, Jasmine, and WFIRST,
][]{Theia,Jasmine,WFIRST}. Nevertheless, in \Fig{fig:vlos} we show how
a large spectroscopic survey \citep[e.g. APOGEE and
  WEAVE,][]{APOGEE,WEAVE} \emph{alone} (i.e. without supplementary
information on proper motions) can be used to probe the effect of the
breathing mode and disentangle between different models of the
non-axisymmetries in the Milky Way disc \citep[for a similar analysis,
  in the Gaia case, see ][]{Antoja2016}. We consider the differential
line of sight velocity $\vlost$, this time defined as
$\vlost=\vlos-\vlosz$, where
\begin{equation}
  \vlos=\frac{(x-x_0)v_x+(y-y_0) v_y + (z-z_0)v_z}{d},
\end{equation}
$(x_0,y_0,z_0)=(8,0,0)$ are the coordinates of the Sun, and $\vlosz$
is the projected $\avvphiz$ on the line-of-sight. We plot $\vlost$ as
a function of the distance from the Sun $d$, and the Galactic
longitude $l$, for the simulations B, S2, and BS2 (top left, top
right, and bottom left panel respectively in \Fig{fig:vlos}), and the
difference $\Delta\vlost$ between $\vlost$ in the BS2 and B
simulations. To make more realistic the comparison with,
e.g. WEAVE\footnote{Note that since WEAVE is a survey of the Northern
  sky, and only the Galactic longitudes ranging from $l \simeq
  20^\circ$ to $l \simeq 225^\circ$ will be observed.}, we consider
only those particles with Galactic latitude
$|b-2\degr|<1\degr$. Moreover, the distance of the particles $d$ is
then convolved with random errors drawn from Gaussian distributions
with standard deviation $0.1d$, in order to simulate the typical error
of the photometric distance estimate of the red clump stars
\citep[see][]{Monari2014}.

We see from these plots how an accurate choice of the lines-of-sight
of the spectroscopic survey would allow to disentangle between
different models of the Milky Way, as the signal in $\vlost$ is
significant at distances between $1$ and $6\Kpc$ from the Sun, with
peak signals of $\sim\pm 15\kmsec$, in the cases with the bar. The
distance error does not blur the signal, and the differences between
the different models remain recognizable.

In particular, from the bottom right panel of \Fig{fig:vlos} we see
that the difference between the BS2 and B models reaches amplitudes of
$\sim 6-9\kmsec$, larger than the typical error in $\vlost$ of a
survey like WEAVE, hence allowing to disentangle bar-only models from
models including both a bar and spiral arms. Even though these
differences are relatively small, more generally speaking the
non-linear enhancement of the mean vertical velocities in the presence
of spiral arms compared to a bar-only case is a major feature that
will allow in the future to disentangle the respective contribution of
the bar and spirals to the Galactic potential. Indeed, no realistic
parameters in a bar-only model could ever reproduce the vertical mean
motions produced in the BS2 model, without causing much larger and
unrealistic radial motions (see M15). Hence, the bar-only models and
bar+spiral models are not degenerate with each other as long as one
considers both the radial and vertical mean motions. However, in
reality the matters are complicated further by the possible influence
of external perturbers on the dynamics of the outer disc (but probably
not that much in the inner disc).

\section{Discussion and Conclusions}\label{sect:concl}
In this work, by means of three-dimensional test-particle numerical
simulations, we have focused on the kinematics of the stars of the
Galactic disc when it is affected by the coupled gravitational
perturbations of a bar and quasi-static spiral arms of different
pattern speeds.

While these effects are essentially the sum of the effects of the
single perturbers for in-plane motions away from major resonances, our
major finding is that significant non-linear motions appear from the
coupling in the vertical kinematics {\it everywhere} in the
disc. These effects are able to double the amplitude of the vertical
breathing modes generated by spirals alone. In particular, there seems
to be an increase of the `compression' (i.e. of the number of stars
with velocity pointing towards the Galactic plane) on the top of the
arms and of the `rarefaction' (i.e. of the number of stars moving
away from the plane) in the interarm regions.

Looking at the power spectrum of the peculiar line-of-sight velocity,
like in the recent work of \cite{Bovy2015}, does not make these
non-linear effects appear because this method makes use only of the
component of the velocity parallel to the Galactic plane. If confirmed
by further surveys, the large-scale velocity fluctuations observed
with APOGEE are thus indeed predominantly affected by the Galactic
bar, provided the bar is strong enough. But the amplitude of spirals
certainly does not have to be negligible for the power spectrum to
match APOGEE observations. In order to make use of the noticeable
differences in terms of vertical motions, we have shown how it will be
possible for a spectroscopic survey like WEAVE alone to distinguish
between bar-only and bar+spiral models. The way to do this is to use
the true (and not projected) line-of-sight velocity, to have fields of
view slightly inclined with respect to the Galactic plane, and
distributed rather continuously in Galactic longitudes, like it was
suggested also by \cite{Kawata2014,Hunt2015}. The typical distance
error expected for red clump giants does not blur the signal of the
different models. Note that the differences are not solely due to
differences in $\avvz$, but also in $\avvphi$ and $\avvR$.

Finally, in agreement with previous two-dimensional investigations
\citep{MinchevFamaey2010}, we confirmed that the coupling of the bar
and spiral arms enhances the radial migration even when the spiral
amplitude is non-varying, and even when low-order spiral resonances
overlap with the bar. We found a significant fraction of orbits
crossing the OLR  of the bar. This means that the
OLR is not a barrier separating the chemical evolution and churning of
the inner disc from the outer disc \citep{Halle2015}, although in the
absence of significant spiral arms, stars from the inner disc do not
migrate to the outer disc.

In conclusion, the non-linear effects due to the coupling of a bar and
spiral arms of different pattern speeds are significant, and have
to be taken in consideration in future models of the Galaxy. To do
this, theoretical insight of the problem is necessary, in the spirit
of M16, but with two perturbers instead of one, which will be the
topic of a follow-up paper.

\section*{Acknowledgments}
We thank the anonymous referee for insightful comments on this
manuscript and Jo Bovy for useful discussions. This work has been
supported by a postdoctoral grant from the {\it Centre National
  d'Etudes Spatiales} (CNES) for GM. RG acknowledges support through
the DFG Research Centre SFB-881 `The Milky Way System' through project
A1.

\bibliographystyle{mn2e}
\bibliography{mn-jour,barspbib}

\begin{appendix}

\end{appendix}

\label{lastpage}

\end{document}